\def\uw{\bds{w}}
\def\ud{\bds{d}}

\def\Kd{\ton{\kA\bds\cdot\ud}}
\def\Kw{\ton{\kA\bds\cdot\uw}}

\def\Kl{\ton{\kA\bds\cdot\ul}}
\def\Km{\ton{\kA\bds\cdot\um}}
\def\Kn{\ton{\kA\bds\cdot\mathbf{\hat{\textrm{N}}}}}

\def\ul{{\bds{\hat{l}}}}
\def\um{{\bds{\hat{m}}}}

\def\kA{{\bds{\hat{S}}}}

\def\nk{n_{\rm b}}

\def\snf{\sin f}
\def\csf{\cos f}
\def\cu{\cos u}
\def\su{\sin u}

\def\rfr#1{Equation~(\ref{#1})}
\def\rfrs#1#2{Equations~(\ref{#1})~to~(\ref{#2})}

\def\dert#1#2{\frac{{{\textrm{d}}}{#1}}{{{\textrm{d}}}{#2}}}
\def\eqi{\begin{equation}}
\def\eqf{\end{equation}}
\def\eqia{\begin{eqnarray}}
\def\eqfa{\end{eqnarray}}

\def\rp#1#2{{#1\over#2}}
\def\lb#1{\label{#1}}

\def\bds#1{\boldsymbol{#1}}
\def\co{\cos\omega}
\def\so{\sin\omega}

\def\cO{\cos\Omega}
\def\sO{\sin\Omega}

\def\cI{\cos I}
\def\sI{\sin I}

\def\ton#1{\left(#1\right)}
\def\qua#1{\left[#1\right]}
\def\grf#1{\left\{#1\right\}}

\documentclass{aastex}

\usepackage{hyperref}
\usepackage[utf8x]{inputenx}
\usepackage[LGR,T1,T5]{fontenc}
\usepackage{float}
\usepackage{amsmath,textgreek,w-greek,wasysym}
\usepackage{amscd,lineno}
\usepackage{amssymb}
\usepackage{graphicx,epsfig}
\usepackage{txfonts}
\usepackage[polutonikogreek,english]{babel}
\usepackage{xr-hyper}
\RequirePackage{color}
\newcommand{\ihat}{\boldsymbol{\hat{\textbf{\i}}}}

\newcommand{\emaila}{lorenzo.iorio@libero.it}
\newcommand{\grk}[1]{\selectlanguage{polutonikogreek}
#1\selectlanguage{english}}
\allowdisplaybreaks[1]

\begin{document}

\title{Analytically  calculated  post-Keplerian range and range-rate perturbations: the solar Lense-Thirring effect and BepiColombo}

\shortauthors{L. Iorio}

\author{Lorenzo Iorio\altaffilmark{1} }
\affil{Ministry of Education, Universities and Research. Department for Education.
\\ Permanent address for correspondence: Viale Unit\`{a} di Italia 68, 70125, Bari (BA),
Italy}

\email{\emaila}

\begin{abstract}
We analytically calculate the time series for the perturbations $\Delta\rho\ton{t},~\Delta\dot\rho\ton{t}$ induced by a general disturbing acceleration $\bds A$ on the mutual range $\rho$ and range-rate $\dot\rho$ of two test particles $\textrm{A},~\textrm{B}$ orbiting the same spinning body.
%of mass $M$, angular momentum $\bds S$, equatorial radius $\mathcal{R}$ and oblateness $J_2$.%
We apply it to the general relativistic Lense-Thirring effect, due to the primary's spin $\bds S$, and the classical perturbation arising from its quadrupole mass moment $J_2$ for arbitrary orbital geometries and orientation of the source's symmetry axis $\bds{\hat{S}}$. The Earth-Mercury range and range-rate are nominally affected by the Sun's gravitomagnetic field to the $10~\textrm{m},~10^{-3}~\textrm{cm~s}^{-1}$ level, respectively, during the extended phase  (2026-2028) of the forthcoming BepiColombo mission to Mercury  whose expected tracking accuracy is of the order of $\simeq 0.1~\textrm{m},~2\times 10^{-4}~\textrm{cm~s}^{-1}$. The competing signatures due to the solar quadrupole $J_2^\odot$, if modelled at the $\sigma_{J_2^\odot}\simeq 10^{-9}$ level of the latest planetary ephemerides INPOP17a, are nearly 10 times smaller than the relativistic gravitomagnetic effects. The position and velocity vectors $\mathbf{r},~\mathbf{v}$ of Mercury and Earth are changed by the solar Lense-Thirring effect by about $10~\textrm{m},~1.5~\textrm{m}$ and $10^{-3}~\textrm{cm~s}^{-1},~10^{-5}~\textrm{cm~s}^{-1}$, respectively, over 2 yr; neglecting such shifts may have an impact on long-term integrations of the inner solar system dynamics over $\sim\textrm{Gyr}$ timescales.
\end{abstract}

keywords{
gravitation--celestial mechanics--Sun: rotation--space veichles
}
%keywords{Experimental studies of gravity; Experimental tests of gravitational theories; Lunar, planetary, and deep-space probes}
\section{Introduction}
Let us consider a pair of test particles $\textrm{A},~\textrm{B}$ following two generally different elliptical paths around the same spinning, oblate massive body of mass $M$, equatorial radius $\mathcal{R}$, quadrupole mass moment $J_2$ and angular momentum $\bds S$, whose symmetry axis is arbitrarily oriented in space; they could typically be two planets orbiting a star, two man-made spacecraft circling a planet or a satellite \citep{1969JGR....74.5295W} like the GRACE \citep{2004GeoRL..31.9607T} and GRAIL \citep{2013SSRv..178....3Z} space missions, or a subset of  constellations of more than two spacecraft like, e.g., the planned eLISA \citep{2013GWN.....6....4A}. The two-body range $\rho$, i.e. the mutual distance $\rho$ between A and B, and its temporal rate of change $\dot\rho$, i.e. the range-rate, contain a wealth of information about the evolution of the system itself and the properties of the common primary's gravitational field in which $\textrm{A},~\textrm{B}$ move. Think about, e.g., the past and future long-term evolution of our solar system which has been-and will perhaps be-studded with collisions among planetesimals and planets themselves \citep{2001Natur.412..708C,2009Natur.459..817L}, or, to much shorter timescales, the accurate mapping of the Earth's gravity field performed by GRACE and its planned follow-on GRACE-FO \citep{2012JGeod..86..319L}. At intermediate time-scales, geocentric range and range-rate measurements to spacecraft orbiting some major bodies of our solar system are able to improve, among other things, their orbit determination \citep{2011Icar..211..401K,2014A&A...561A.115V,2015IAUGA..2244873F}. As a consequence, intersatellite and interplanetary tracking through electromagnetic waves of different frequencies have reached an impressive accuracy level \citep{2003SSRv..108..377B,2009AcAau..65.1571K, 2012JGeod..86.1083S,2009AcAau..65..666I}.
One of the main application of accurate ranging is testing the general theory of relativity, whose current status was recently overviewed by, e.g., \citet{2016Univ....2...23D}.
In this paper, we analytically calculate the mutual two-body range and range-rate perturbations $\Delta\rho,~\Delta\dot\rho$ induced by the general relativistic gravitomagnetic\footnote{It is generated by mass-energy currents encoded in the $g_{0j},~j=1,2,3$ components of the spacetime metric tensor $g_{\mu\nu},~\mu,\nu=0,1,2,3$ \citep{1986hmac.book..103T}. It is believed to play important roles in astrophysical scenarios involving spinning black holes \citep{1988nznf.conf..573T}. At present, its main experimental confirmation relies upon the measurement performed by the GP-B probe in the Earth's gravitational field, accurate to $19$ per cent \citep{2011PhRvL.106v1101E}; for other performed, ongoing and proposed attempts involving the analysis of the orbital motions of natural and artificial test particles around spinning planets and stars, see, for example, \citep{2011Ap&SS.331..351I,2013NuPhS.243..180C,2013CEJPh..11..531R}.} field of the common primary for an arbitrary orientation of its angular momentum and without making a-priori simplifying assumptions on the eccentricity of both the orbiting test particles. This choice is motivated by reasons of generality and in view of the fact that the spatial orientation of the angular momentum of astronomical bodies which could be considered as potentially interesting sources of testable gravitomagnetic fields is not always known with sufficient accuracy. Furthermore, the coordinate systems used in several practical astronomical data analyses do not generally adopt the primary's equator as their fundamental plane. Moreover, over multiyear data analyses, the angular momentum of the source of the gravitomagnetic field often changes its position with respect to that originally occupied at the reference epoch of the coordinate system used because of a variety of more or less accurately known physical phenomena; also such time-dependent evolution is necessarily known with a given level of uncertainty. Our general approach is fully able to cope with such issues allowing for suitably designing dedicated scenarios and performing sensitivity analyses in view of, e.g., forthcoming space-based missions like BepiColombo at Mercury \citep{2007SSRv..132..611B,2016Univ....2...21S}. This is important either if one is considering to put directly to the test the Lense-Thirring effect via range and range-rate measurements, and if, conversely, one trusts general relativity also as far as the gravitomagnetic field is concerned and uses it as a potential tool to measure the angular momentum of an astronomical body. Even if one is interested in other dynamical effects, like, e.g., characterizing the Newtonian multipolar structure of the gravitational potential of its source, the Lense-Thirring effect must be taken into account to avoid a priori biased results, as it would be the case for the Sun's oblateness \citep{2014IPNPR.196C...1F}.
For the sake of completeness, we deal also with the Schwarzschild-type gravitoelectric range and range-rate perturbations. Almost all the results required to build them can be already found in the literature except the short-term variation of the mean anomaly, which is correctly calculated in the present paper.
It is important to point out that, in view of its generality and lack of a priori restrictions on the orbital geometries involved, our strategy can be straightforward extended to any other perturbing acceleration induced, e.g., by modified models of gravity as well.

The plan of the paper is as follows. In Section~\ref{teoria}, we outline our general  approach to analytically calculate the perturbations induced by any small post-Keplerian disturbing acceleration on the mutual range and range-rate of a pair of test particles orbiting a common primary. It is applied to the Schwarzschild-type acceleration in Section~\ref{GEpert} and to the Lense-Thirring effect in Section~\ref{LTpert}, where exact analytical expressions for the building blocks of the range and range-rate gravitomagnetic shifts are explicitly displayed.
Section~\ref{J2pert} deals with the range and range-rate perturbations due to the first even zonal harmonic $J_2$ of the multipolar expansion of the primary's gravitational field since it is a major source of systematic bias of dynamical origin to be accounted for in any realistic preliminary sensitivity analysis. In view of their extreme cumbersomeness in the case of arbitrary spatial orientation of the primary's spin axis, it is not possible to explicitly show the analytical results obtained by applying the general method of Section~\ref{teoria} to this specific disturbance. The shifts treated in Sections~\ref{LTpert}~to~\ref{J2pert} are due only to the orbital motions of $\textrm{A},~\textrm{B}$ in the field of their primary. We will show that the effects of the propagation of the electromagnetic waves in its gravitomagnetically deformed spacetime are negligible in the specific scenario treated in Section~\ref{bepi} which deals with the geocentric Hermean\footnote{From \grk{<Erm~hs} (`Hermes'), corresponding to the ancient Roman deity Mercury} range and range-rate during the planned two-years extended mission of BepiColombo orbiting Mercury. In Section~\ref{caos}, we apply the results of Section~\ref{LTpert} to evaluate the corrections to the Hermean and terrestrial state vectors in view of possible consequences of neglecting the Sun's gravitomagnetic field in the so-far performed long-term numerical integrations of the solar system dynamics over future eons. Our finding and conclusions are summarized in Section~\ref{fine}.
\section{General calculational method of the range and range-rate perturbations}\lb{teoria}
According to \citep{2002JGeod..76..169C}, the perturbation $\Delta\rho$ of the mutual range $\rho$ of two test particles $\textrm{A},~\textrm{B}$ orbiting the same central body
%of mass $M$, equatorial radius $\mathcal{R}$, quadrupole mass moment $J_2$ and angular momentum $\bds S$%
is
\eqi
\Delta\rho = \ton{\mathbf{\Delta\textrm{r}}_\textrm{A}-\mathbf{\Delta\textrm{r}}_\textrm{B}}\bds\cdot\mathbf{\hat{\rho}}, \lb{range}
\eqf
where
\eqi
\mathbf{\hat{\rho}} = \rp{\ton{\mathbf{r}_\textrm{A}-\mathbf{r}_\textrm{B}}}{\rho},\lb{rho}
\eqf
and
\eqi
\rho^2=\ton{\mathbf{r}_\textrm{A}-\mathbf{r}_\textrm{B}}\bds\cdot\ton{\mathbf{r}_\textrm{A}-\mathbf{r}_\textrm{B}}.\lb{rho2}
\eqf
In turn, the perturbation $\mathbf{\Delta\mathrm{r}}$ experienced by the position vector $\mathbf{r}$ of any of the two bodies $\textrm{A},~\textrm{B}$ is
\eqi\mathbf{\Delta\mathrm{r}}=\Delta R~\mathbf{\hat{R}}+\Delta T~\mathbf{\hat{T}}+\Delta N~\mathbf{\hat{N}},\lb{Derre}\eqf
In \rfr{Derre}, the instantaneous radial, transverse and normal perturbations $\Delta R,~\Delta T,~\Delta N$ of the position vector $\mathbf{r}$, expressed in terms of the osculating semimajor axis $a$, eccentricity $e$, inclination $I$, longitude of the ascending node $\Omega$, argument of pericenter $\omega$, mean anomaly $\mathcal{M}$, are \citep{1993CeMDA..55..209C}
\begin{align}
\Delta R\ton{f} \lb{rR}&= \rp{r\ton{f}}{a}\Delta a\ton{f} -a\cos f\Delta e\ton{f} +\rp{ae\sin f}{\sqrt{1-e^2}}\Delta\mathcal{M}\ton{f}, \\ \nonumber \\
\Delta T\ton{f} \lb{rT}&= a\sin f\qua{1 + \rp{r\ton{f}}{p}}\Delta e\ton{f} + r\ton{f}\qua{\cI\Delta\Omega\ton{f}+\Delta\omega\ton{f}} +\rp{a^2}{r\ton{f}}\sqrt{1-e^2}\Delta\mathcal{M}\ton{f}, \\ \nonumber \\
\Delta N\ton{f} \lb{rN}&= r\ton{f}\qua{\sin u~\Delta I\ton{f} -\sI\cos u~\Delta\Omega\ton{f}}.
\end{align}
In \rfrs{rR}{rN}, $f$ is the true anomaly, assumed as fast variable encoding the time dependence (see \rfr{fMt}), and $u=\omega+f$ is the argument of latitude.
The radial unit vector $\mathbf{\hat{R}}$ entering \rfr{Derre} can be expressed as \citep{1991ercm.book.....B}
\eqi
\mathbf{\hat{R}} = \ul\cos u+\um\sin u,
\eqf
where the unit vectors $\ul,~\um$ are defined as  \citep{1991ercm.book.....B}
\begin{align}
\ul &=\cO~\ihat+\sO~\bds{\hat{\j}}, \\ \nonumber\\
\um &=-\cI\sO~\ihat+\cI\cO~\bds{\hat{\j}}+\sI~\mathbf{\hat{k}}.
\end{align}
The vector $\ul$ is directed along the line of the nodes toward the ascending node, while $\um$ is directed transversely to the line of the nodes in the orbital plane; $\ihat,~\bds{\hat{\j}},~\mathbf{\hat{k}}$ are the usual unit vectors spanning the reference $x,~y,~z$ axes of the coordinate system adopted.
The normal unit vector $\mathbf{\hat{N}}$ in \rfr{Derre} is \citep{1991ercm.book.....B}
\eqi\mathbf{\hat{N}}=\sI\sO~\ihat-\sI\cO~\bds{\hat{\j}}+\cI~\mathbf{\hat{k}}\eqf
Thus, the transverse unit vector $\mathbf{\hat{T}}$ appearing in \rfr{Derre} can straightforwardly be obtained as \citep{1991ercm.book.....B}
\eqi
\mathbf{\hat{T}}=\mathbf{\hat{N}}\bds\times\mathbf{\hat{R}}.
\eqf
Both \rfrs{range}{rho2} and \rfrs{rR}{rN} must be evaluated onto the unperturbed Keplerian ellipse
\eqi
r=\rp{a\ton{1-e^2}}{1+e\cos f},\lb{kepell}
\eqf
assumed as reference orbit; thus the position vector of any of $\textrm{A},~\textrm{B}$, entering \rfrs{rho}{rho2}, is
\eqi
\mathbf{r} = r~\mathbf{\hat{R}},
\eqf
where $r$ is given by \rfr{kepell}.
Furthermore, the instantaneous shifts $\Delta\kappa\ton{f},~\kappa=a,~e,~I,~\Omega,~\omega$ entering \rfr{Derre} are to be calculated as
\eqi
\Delta\kappa\ton{f}=\int_{f_0}^f\dert\kappa{t}\dert{t}{f^{'}}df^{'},~\kappa=a,~e,~I,~\Omega,~\omega,
\eqf
by taking $d\kappa/dt,~\kappa=a,~e,~I,~\Omega,~\omega$ from the right-hand-sides of the usual Gauss equations for the variation of the elements \citep{1991ercm.book.....B,Nobilibook87,1989racm.book.....S,2003ASSL..293.....B}, evaluated onto \rfr{kepell}, and with
\eqi
\dert{t}{f} = \rp{\ton{1-e^2}^{3/2}}{\nk\ton{1+e\cos f}^2},
\eqf
in which $\nk=\sqrt{GMa^{-3}}$ is the Keplerian mean motion; $G$ is the Newtonian constant of gravitation.
The instantaneous change $\Delta\mathcal{M}\ton{f}$ of the mean anomaly $\mathcal{M}=\nk\ton{t-t_p}$, where $t_p$ is the time of passage at pericenter, must be evaluated as
\eqi
\Delta\mathcal{M}\ton{f}=\Delta\eta\ton{f}+\int_{t_0}^t\Delta\nk\ton{t^{'}}dt^{'},\lb{DeltaM}
\eqf
where $\Delta\eta\ton{f}$ is the instantaneous change of the mean anomaly at epoch $\eta$, worked out with the corresponding Gauss equation \citep{1991ercm.book.....B,Nobilibook87,1989racm.book.....S,2003ASSL..293.....B}, and
\eqi
\int_{t_0}^t\Delta\nk\ton{t^{'}}dt^{'}=-\rp{3}{2}\rp{\nk}{a}\int_{f_0}^f\Delta a\ton{f^{'}}\rp{dt}{df^{'}}df^{'},\lb{IntDn}
\eqf
Depending on the specific perturbation at hand, the explicit calculation of \rfr{IntDn} may turn out rather cumbersome.

It is possible to analytically calculate also the shift $\Delta\dot\rho$ experienced by the range-rate $\dot\rho$ as follows \citep{2002JGeod..76..169C}
\eqi
\Delta\dot\rho =  \ton{\mathbf{\Delta\textrm{v}}_\textrm{A}-\mathbf{\Delta\textrm{v}}_\textrm{B}}\bds\cdot\mathbf{\hat{\rho}}+
\ton{\mathbf{\Delta\textrm{r}}_\textrm{A}-\mathbf{\Delta\textrm{r}}_\textrm{B}}\bds\cdot{\mathbf{\hat{\rho}}}_{\textrm{v}},\lb{Drrange}
\eqf
where
\eqi
{\mathbf{\hat{\rho}}}_{\textrm{v}}=\rp{\ton{\mathbf{\textrm{v}}_\textrm{A}-\mathbf{\textrm{v}}_\textrm{B}}-\dot\rho\mathbf{\hat{\rho}}}{\rho},
\lb{vvr}\eqf
with
\eqi
\dot\rho =  \ton{\mathbf{\textrm{v}}_\textrm{A}-\mathbf{\textrm{v}}_\textrm{B}}\bds\cdot\mathbf{\hat{\rho}}.\lb{dotrho}
\eqf
The perturbation $\mathbf{\Delta\mathrm{v}}$ of the velocity vector $\mathbf{v}$ of any of the two bodies $\textrm{A},~\textrm{B}$ can be written as
\eqi
\mathbf{\Delta\mathrm{v}}=\Delta\textrm{v}_R~\mathbf{\hat{R}}+\Delta \textrm{v}_T~\mathbf{\hat{T}}+\Delta \textrm{v}_N~\mathbf{\hat{N}}.\lb{Dvel}
\eqf
In \rfr{Dvel}, the instantaneous radial, transverse and normal perturbations $\Delta\textrm{v}_R,~\Delta\textrm{v}_T,~\Delta\textrm{v}_N$ of the position vector $\mathbf{v}$, expressed in terms of the osculating orbital elements, are \citep{1993CeMDA..55..209C}
\begin{align}
\Delta \mathrm{v}_R\ton{f} \lb{vR} \nonumber & = -\rp{\nk a \sin f}{\sqrt{1-e^2}}\qua{\rp{e}{2a}\Delta a\ton{f} + \rp{a}{r\ton{f}}\Delta e\ton{f} } -\rp{\nk a^3}{r^2\ton{f}}\Delta{\mathcal{M}}\ton{f}  - \\ \nonumber \\
&-\rp{\nk a^2}{r\ton{f}}\sqrt{1-e^2}\qua{\cI\Delta\Omega\ton{f} + \Delta\omega\ton{f}}, \\ \nonumber \\
\Delta \mathrm{v}_T\ton{f} \lb{vT} & = -\rp{\nk a\sqrt{1-e^2}}{2r\ton{f}}\Delta a\ton{f} + \rp{\nk a \ton{e+\cos f} }{\ton{1-e^2}^{3/2}}\Delta e\ton{f} +\rp{\nk a e \sin f}{\sqrt{1-e^2}}\qua{\cI\Delta\Omega\ton{f} + \Delta\omega\ton{f}}, \\ \nonumber \\
\Delta \mathrm{v}_N\ton{f} \lb{vN} & = \rp{\nk a}{\sqrt{1-e^2}}\qua{\ton{\cos u + e \cos\omega}\Delta I\ton{f} + \ton{\sin u + e \so}\sin I\Delta\Omega\ton{f}  }.
\end{align}
Since, also in this case, the Keplerian ellipse has to be adopted as unperturbed reference trajectory,
\eqi
\mathbf{v} = \sqrt{\rp{GM}{a\ton{1-e^2}}}\qua{-\bds{\hat{P}}\sin f + \bds{\hat{Q}}\ton{e+\cos f}}\lb{Kepvel}
\eqf
must be used in \rfrs{vvr}{dotrho}. The unit vectors $\bds{\hat{P}},~\bds{\hat{Q}}$ in \rfr{Kepvel} are defined as \citep{1991ercm.book.....B}
\begin{align}
\bds{\hat{P}} = \ul\cos\omega+\um\sin\omega, \\ \nonumber \\
\bds{\hat{Q}} = -\ul\sin\omega+\um\cos\omega;
\end{align}
while $\bds{\hat{P}}$ is directed along the line of the apsides toward the pericenter, $\bds{\hat{Q}}$ lies transversely to the line of the apsides in
the orbital plane. Incidentally, it turns out that also the position vector $\mathbf{r}$ can be expressed in terms of $\bds{\hat{P}},~\bds{\hat{Q}}$ as \citep{1991ercm.book.....B}
\eqi
\mathbf{r} = r\ton{\bds{\hat{P}}\cos f + \bds{\hat{Q}}\sin f}.
\eqf
\section{The general relativistic range and range-rate perturbations}
\subsection{The gravitoelectric Schwarschild-like effect}\lb{GEpert}
As far as the post-Newtonian  gravitoelectric\footnote{In our solar system, it induces the formerly anomalous, time-honored perihelion precession of Mercury of $\dot\omega^\textrm{GE}_{\mercury} = 42.98~\textrm{arcsec~cty}^{-1}$ \citep{1986Natur.320...39N}.} acceleration due to the static part of the primary's field is concerned, most of the building blocks required to calculate its range and range-rate perturbations are available in the literature. Indeed, the instantaneous variations $\Delta\kappa^\textrm{GE}\ton{f},~\kappa=a,~e,~I,~\Omega,~\omega$ can be found in Equations~(A2.78b)~to~(A2.78d) of \citet[p.~178]{1989racm.book.....S}.
The correct calculation of the variation of the mean anomaly requires more care. According to \rfrs{DeltaM}{IntDn}, its instantaneous change turns out to be
\begin{equation}
\Delta\mathcal{M}^\textrm{GE}\ton{f} = C_1\grf{\arctan\qua{\rp{\ton{-1+e}\tan\ton{\rp{f}{2}}}{\sqrt{1-e^2}}}  -\arctan\qua{\rp{\ton{-1+e}\tan\ton{\rp{f_0}{2}}}{\sqrt{1-e^2}}}} + C_2,\label{DMf}
\end{equation}
with
\begin{align}
C_1 \nonumber \lb{C1}& = \rp{G\ton{M+m}}{4c^2 a\ton{1-e^2}^2}
\ton{72 + e^2 \ton{84 - 76 \nu} + 4 e^4 \ton{6 - 7 \nu} - 16 \nu + \right.\\ \nonumber \\
&+\left. 3 e\ton{\ton{56 + e^2 \ton{24 - 31 \nu} - 24 \nu} \cos f_0 + e \ton{4 \ton{5 - 4 \nu} \cos 2 f_0 - e \nu \cos 3 f_0}}}, \\ \nonumber \\
C_2
\nonumber \lb{C2}& =
\rp{G\ton{M+m}}{16c^2ae\ton{1-e^2}^2\ton{1+e\cos f}}
\ton{4 e \uppi \ton{1 + e \cos f} \ton{8 \ton{-9 + 2 \nu} + 4 e^4 \ton{-6 + 7 \nu} + \right.\right.\\ \nonumber \\
\nonumber &+ \left.\left. e^2 \ton{-84 + 76 \nu} + 3 e \ton{\ton{8 \ton{-7 + 3 \nu} + e^2 \ton{-24 + 31 \nu}} \cos f_0 + \right.\right.\right.\\ \nonumber \\
\nonumber &+ \left.\left.\left. e \ton{4 \ton{-5 + 4 \nu} \cos 2f_0 + e \nu \cos 3f_0}}} + 2 \sqrt{1-e^2} \ton{\ton{e^4 \ton{66 - 83 \nu} + e^2 \ton{90 - 29 \nu} -\right.\right.\right.\\ \nonumber \\
\nonumber &- \left.\left.\left.  8 \ton{-3 + \nu} + 3 e^3 \ton{\ton{56 + e^2 \ton{24 - 31 \nu} - 24 \nu} \cos f_0 + \right.\right.\right.\right.\\ \nonumber \\
\nonumber &+ \left.\left.\left.\left. e \ton{4 \ton{5 - 4 \nu} \cos 2f_0 - e \nu \cos 3f_0}}}\sin f + \rp{e}{2} \ton{-1 + e^2} \ton{\ton{14 e^2 \ton{-4 + 3 \nu} + \right.\right.\right.\right.\\ \nonumber \\
\nonumber &+ \left.\left.\left.\left. 8 \ton{-8 + 5 \nu}} \sin 2f + e \ton{2 \ton{-10 + 9 \nu} \sin 3f + e \nu \sin 4f}} + \right.\right.\\ \nonumber \\
\nonumber &+\left.\left. \ton{8 \ton{-3 + \nu} + e^2 \ton{-80 + 21 \nu} + 2 e^4 \ton{-8 + 23 \nu}} \ton{1 + e \cos f} \sin f_0 + \right.\right.\\ \nonumber \\
\nonumber &+\left.\left. 2 e \ton{-10 + 8 \nu + e^2 \ton{-20 + 13 \nu}} \ton{1 + e \cos f} \sin 2f_0 + \right.\right.\\ \nonumber \\
&+\left.\left. e^2 \ton{1 + 2 e^2} \nu \ton{1 + e \cos f} \sin 3f_0}}.
\end{align}
In \rfrs{C1}{C2}, the dimensionless mass parameter $\nu= m M \ton{M+m}^{-2}$, where $m$ is the mass of any of the two objects $\textrm{A,~B}$, vanishes in the test particle limit.
Equations~(A2.78e)~to~(A2.78f) of \citet[p.~178]{1989racm.book.....S} allow to obtain the instantaneous shift of the mean anomaly in terms of the three anomalies $f,~E,~\mathcal{M}$; instead, only the true anomaly $f$ enters our \rfr{DMf}. See also Equations~(3.1.102)~to~(3.1.107) of \citet[p.~93]{1991ercm.book.....B}.
Let us remark that \rfr{DMf}
%, which is a novel result amending several incorrect formulas existing in the literature,
is exact in $e$.
\subsection{The gravitomagnetic Lense-Thirring effect}\lb{LTpert}
The Lense-Thirring acceleration ${\bds A}_\textrm{LT}$, written in harmonic, post-Newtonian coordinates, can be found in several references; see, e.g., \citet{1991ercm.book.....B,1989racm.book.....S,2010ITN....36....1P}. Its radial, transverse and normal components, for a generic orientation of the spin axis unit vector $\kA$, i.e. for $\kA\neq\mathbf{\hat{k}}$, can be retrieved, e.g., in \citet{2017EPJC...77..439I}; they are
\begin{align}
A_R^\textrm{LT} \lb{ARLT} & = \rp{2G\nk \ton{1+e\csf}^4\ton{\bds\kA\bds\cdot\mathbf{\hat{N}}}}{c^2 a^2\ton{1-e^2}^{7/2}}, \\ \nonumber \\
A_T^\textrm{LT} \lb{ATLT} & = -\rp{2eG\nk \ton{1+e\csf}^3\snf\ton{\kA\bds\cdot\mathbf{\hat{N}}}}{c^2 a^2\ton{1-e^2}^{7/2}}, \\ \nonumber \\
A_N^\textrm{LT} \nonumber \lb{ANLT} & = -\rp{2G\nk\ton{1+e\csf}^3}{c^2 a^2 \ton{1-e^2}^{7/2}}\kA\bds\cdot\grf{\qua{e\co -\ton{2+3e\csf}\cu}\ul -\right.\\ \nonumber \\
&-\left. \rp{1}{2}\qua{e\so +4\su + 3e\sin\ton{\omega+2f}  }\um  },
\end{align}
where $c$ is the speed of light.
Thus,
the shifts $\Delta R_\textrm{LT},~\Delta T_\textrm{LT},~\Delta N_\textrm{LT}$ of the radial, transverse and normal components of the position vector $\mathbf{r}$ turn out to be
\begin{align}
\Delta R_\textrm{LT} & = \rp{2 G S\Kn  \qua{1 - \cos\ton{f - f_0}}}{c^2 a^2 \nk \sqrt{1 - e^2}},\\ \nonumber \\
\Delta T_\textrm{LT} & =\rp{G S\Kn  \grf{ 2 e \qua{-1 + \cos\ton{f - f_0}} \sin f + 4 \qua{-f + f_0 + \sin\ton{f - f_0}} }}{c^2 a^2 \nk \sqrt{1 - e^2}\ton{1 + e \cos f}},\\ \nonumber \\
\Delta N_\textrm{LT} & =\rp{G S\Kw}{ c^2 a^2 \nk \sqrt{1 - e^2}\ton{1 + e \cos f}},
\end{align}
where
\begin{align}
\bds{w} \nonumber &= 2\grf{\ton{1+e\cos f_0}\cos u_0\sin\ton{f-f_0}+\qua{f_0- f + e\ton{\sin f_0-\sin f}}\cos u  }\um + \\ \nonumber \\
\nonumber & + \grf{  -\cos\ton{f-\omega - 2 f_0} + \cos u + 2 e \qua{\cos\omega + 2 \cos\ton{u + f_0} + \cos\ton{2 f_0 + \omega}} \sin^2\ton{\rp{f - f_0}{2}} + \right.\\ \nonumber \\
&\left.+ 2 \ton{f - f_0} \sin u   }\ul\lb{wvec}.
\end{align}
From \rfr{wvec}, it can be noted that $\bds w$ is not a unit vector.
The shifts $\Delta\textrm{v}^\textrm{LT}_R,~\Delta \textrm{v}^\textrm{LT}_T,~\Delta \textrm{v}^\textrm{LT}_N$ of the radial, transverse and normal components of the velocity vector $\mathbf{v}$ are
\begin{align}
\Delta\textrm{v}^\textrm{LT}_R  & = \rp{2GS\Kn\ton{1 + e \cos f} \qua{2 \ton{f -  f_0} + e \ton{\sin f - \sin f_0} - \sin\ton{f - f_0}}}{c^2 a^2\ton{1-e^2}^2}, \\ \nonumber \\
\Delta\textrm{v}^\textrm{LT}_T \nonumber & = \rp{GS\Kn}{c^2 a^2\ton{1-e^2}^2}\grf{-2 - e^2 + 2 \ton{e + \cos f} \cos f_0 +\right.\\ \nonumber \\
&+\left. e \qua{-2 \cos f + e \cos 2 f + 4 \ton{-f + f_0} \sin f} + 2 \ton{1 + e^2} \sin f \sin f_0},\\ \nonumber \\
\Delta\textrm{v}^\textrm{LT}_N& = \rp{GS\Kd}{2c^2 a^2\ton{1-e^2}^2},
\end{align}
where
\begin{align}
\bds{d} \nonumber & =
\grf{ 2\qua{-\cos u + \cos\ton{f - 2 f_0 - \omega} + 2 \ton{f - f_0} \sin u}  + \right.\\ \nonumber \\
\nonumber & + \left. e\qua{  -\cos\ton{f_0 - u} - 6 \cos 2 u + 3 \cos\ton{f_0 + u} + \cos\ton{f - u_0} + \right.\right.\\ \nonumber \\
\nonumber &+\left.\left. \cos\ton{f - 3 f_0 - \omega} + 2 \cos\ton{2 f_0 + \omega} + 4 \ton{f - f_0} \sin\omega  } + \right.\\ \nonumber \\
\nonumber &+\left. e^2\qua{-4 \cos u - \cos 3 u + 4 \cos u_0 + \cos\ton{f - \omega} - 2 \sin 2 f_0 \sin u_0}  }\um, \\ \nonumber \\
\nonumber &+\grf{  4 \ton{f - f_0} \cos u + 2 \qua{\sin u + \sin\ton{f - 2 f_0 - \omega}} + \right.\\ \nonumber \\
\nonumber &+\left. e\qua{ 4 \ton{f - f_0} \cos\omega - \sin\ton{f_0 - u} + 6 \sin 2 u - 3 \sin\ton{f_0 + u} +\sin\ton{f - u_0} + \right.\right.\\ \nonumber \\
\nonumber &+\left.\left. \sin\ton{f - 3 f_0 - \omega} - 2 \sin\ton{2 f_0 + \omega} } + \right.\\ \nonumber \\
&+\left. e^2\qua{-2 \cos u_0 \sin 2 f_0 + 4 \sin u + \sin 3 u - 4 \sin u_0 +
 \sin\ton{f - \omega}}    }\ul.\lb{dvec}
\end{align}
According to \rfr{dvec}, $\bds d$ is not a  unit vector.
\section{The oblateness-induced range and range-rate perturbations}\lb{J2pert}
If the main target is testing some predictions of general relativity, one of the major sources of systematic uncertainties is represented by the departures from spherical symmetry of the primary. They are usually modeled as an expansion in multipoles of its gravitational potential. A class of such multipolar coefficients cause long-term Newtonian orbital perturbations whose nominal size is usually orders of magnitude larger than the non-Newtonian effects of interest. Conversely, the determination of the multipoles themselves can be one of the main scientific goals of a given mission; in this case, it is the Lense-Thirring effect that, if not explicitly modeled, can bias the outcome of the experiment, as recently recognized for the Sun by \citet{2014IPNPR.196C...1F} on the basis of \citet{2011Ap&SS.331..351I}.

As far as the disturbing acceleration induced by the quadrupole mass moment $J_2$ of the primary is concerned, its radial, transverse and normal components  for an arbitrary orientation of the unit vector $\kA$ of its symmetry axis are \citet{2017EPJC...77..439I}
\begin{align}
A_{R}^{J_2} \lb{ARJ2}& = \rp{3\mu J_2{\mathcal{R}}^2\ton{1+e\csf}^4}{2a^4\ton{1-e^2}^4}\grf{3\qua{\cu\Kl + \su\Km}^2 - 1}, \\ \nonumber \\
A_{T}^{J_2} \lb{ATJ2}& = -\rp{3\mu J_2{\mathcal{R}}^2\ton{1+e\csf}^4}{a^4\ton{1-e^2}^4}\qua{\cu\Kl + \su\Km}\qua{\cu\Km - \su\Kl}, \\ \nonumber \\
A_{N}^{J_2} \lb{ANJ2}& =-\rp{3\mu J_2{\mathcal{R}}^2\ton{1+e\csf}^4}{a^4\ton{1-e^2}^4}\qua{\cu\Kl + \su\Km}\Kn,
\end{align}
where $\mu=GM$ is the gravitational parameter of the primary.

Without recurring to some a priori approximations about the orbital configuration of the test particle and the orientation of the primary's spin axis, very cumbersome analytical expressions are obtained for $\Delta R_{J_2},~\Delta T_{J_2},~\Delta N_{J_2},~\Delta\textrm{v}_R^{J_2},~\Delta\textrm{v}_T^{J_2},~\Delta \textrm{v}_N^{J_2}$ from \rfrs{ARJ2}{ANJ2}; thus, they cannot be explicitly displayed. In particular, the calculation of \rfr{IntDn} entering the total shift of the mean anomaly in  \rfr{DeltaM} turns out to be particularly unwieldy. They can be conveniently simplified depending on the specific scenario at hand by expanding them in powers of $e$ to the desired level of accuracy.
\section{The BepiColombo range and range-rate perturbations}\lb{bepi}
As a concrete application of our results in Sections~\ref{LTpert}~to~\ref{J2pert}, let us consider the geocentric Hermean range and range-rate during the expected extended  phase (2026 March 14-2028 May 1) of the BepiColombo mission to Mercury \citep{2010P&SS...58....2B,2017EPSC...11..508B}; according to http://sci.esa.int/bepicolombo/47346-fact-sheet/, its launch is currently scheduled in October 2018. It should greatly improve, among other things, the accuracy of the orbital determination of the small rocky planet to the $\sigma_\rho\simeq 0.1~\textrm{m},~\sigma_{\dot\rho}\simeq 2\times 10^{-4}~\textrm{cm~s}^{-1}$ level for the range and the range-rate, respectively \citep{2010IAUS..261..356M}. Here, we will not deal in detail with the gravitoelectric effect of Section~\ref{GEpert} since it has been already treated numerically in the literature \citep{2002PhRvD..66h2001M,2007PhRvD..75b2001A,2010IAUS..261..356M,2016Univ....2...21S,2018Icar..301....9I}.
Recently, more and more extended portions of the data record of the NASA MESSENGER mission to Mercury, ended on April 30, 2015, have started to be analyzed in order to look at the Lense-Thirring effect explicitly by modeling it \citep{2015IAUGA..2227771P,2017AJ....153..121P,2018NatureG}; the first preliminary results evidenced a strong correlation of the gravitomagnetic signature with the solar oblateness, thus limiting the accuracy of a possible direct detection of the relativistic effect to about $\sim 20-25\%$. The perspectives of a direct detection of the solar gravitomagnetic field opened up by the recent advances in the field of the planetary ephemerides were pointed out by the present author more than a decade ago; see, e.g., \citet{2011Ap&SS.331..351I} and references therein.

Before proceeding further, it is advisable to briefly review some basic features of the confrontation between theory and observations in Relativistic Celestial Mechanics (RCM) \citep{2010CeMDA.106..209B,2010SchpJ...510669B,2011rcms.book.....K,2013SoSyR..47..347B}. Such a task implies solving  not only the dynamics of the specific problem at hand, i.e. the equations of motion of the  massive bodies involved, but also the equations of propagation of the electromagnetic waves and the description of the observational procedures (the kinematical part of RCM). Both parts should be investigated in the same coordinates to exclude unphysical coordinate–dependent spurious effects and to present the results in terms of measurable quantities. Indeed, contrary to the Newtonian case, the issue of coordinate–dependent quantities is the main qualitative new feature of RCM. Among the main possibilities to overcome such a problem emerged in the literature over the past decades, in 1991 it was pragmatically adopted by the International Astronomical Union (IAU) the approach of forgetting about the general relativistic arbitrariness in the coordinate conditions, and to use one specific type of coordinates-the harmonic ones-for both the dynamical and the kinematical parts of RCM once for all; see \citet{2003AJ....126.2687S} for the more recent IAU 2000 resolutions updating the earlier IAU 1991 ones. In the case of interplanetary ranging  experiments in the Solar System, it is customary to adopt some realization of a Barycentric Reference System (BRS) along with some suitable Barycentric Coordinate Time (TCB) as global coordinate system. Suitable and well established spacetime coordinate transformations from the specific local Planetocentric Reference Systems (PRSs) of the major bodies involved in the experiment under examination to the BRS allow to avoid the insurgence of spurious terms like in the nowadays outdated problem of the unphysical effects plaguing the Earth-Moon range when calculated in the BRS instead of some realization of a Geocentric Reference System (GRS)\footnote{In analyzing the data provided by the Earth-based Lunar Laser Ranging (LLR) technique, a GRS is more suited than a BRS for the reasons explained in, e.g., \citet{2010CeMDA.106..209B}.} \citep{2010CeMDA.106..209B}. For details of the computation of observables for BepiColombo and the required coordinate transformations, see, e.g., Section 3.1 of \citet{2016Univ....2...21S}.

\begin{figure*}
\centerline{
\begin{tabular}{c}
\epsfxsize= 12 cm\epsfbox{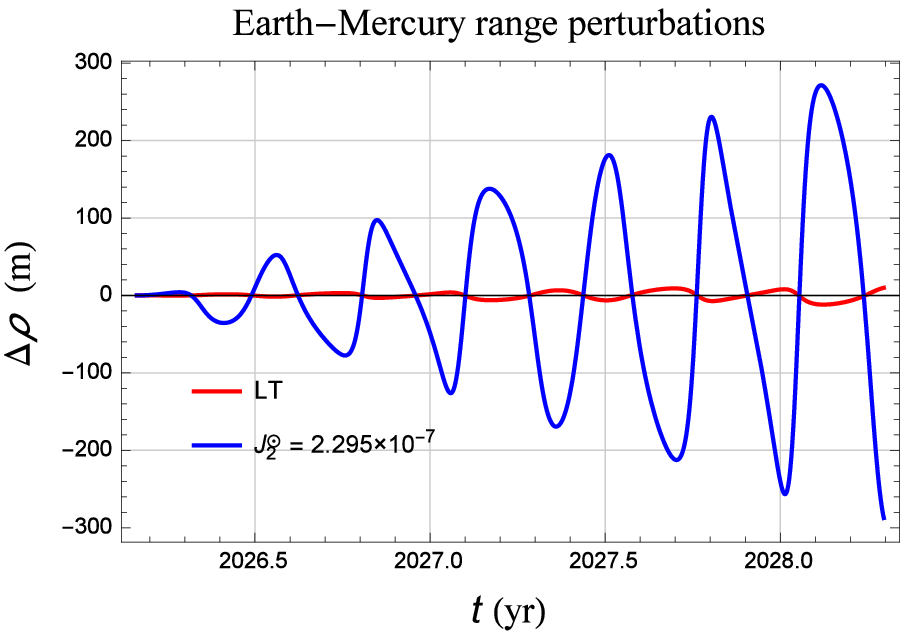}\\
\epsfxsize= 12 cm\epsfbox{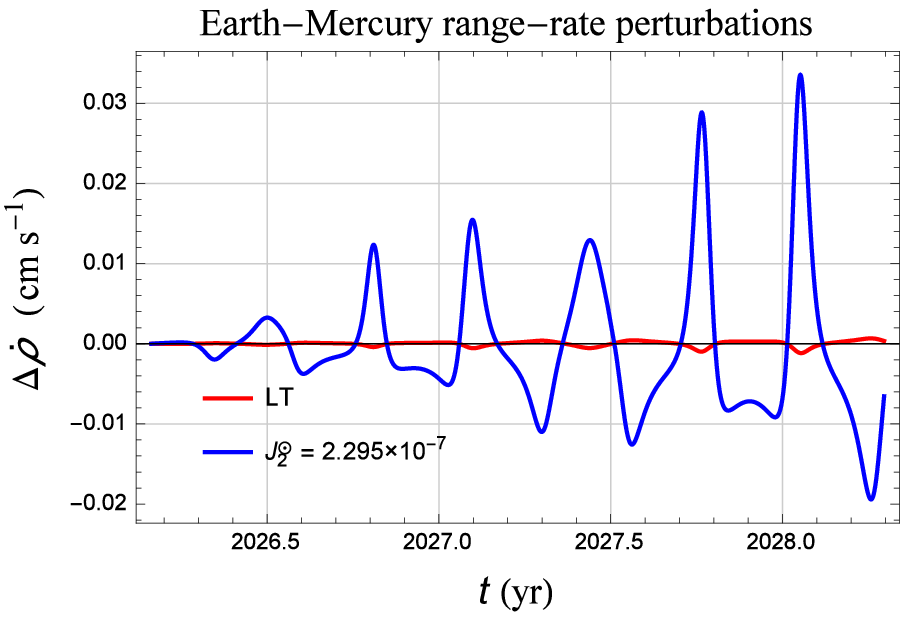}\\
\end{tabular}
}
\caption{Nominal Lense-Thirring and $J_2^\odot$ perturbations of the Earth-Mercury range (in~\textrm{m}) and range-rate (in~$\textrm{cm~s}^{-1}$) during the expected extended mission of Bepi Colombo from 2026 March 14 to 2028 May 1. A coordinate system with the mean ecliptic at the epoch J2000.0 as fundamental reference $\grf{x,~y}$ plane was assumed. The initial values of the Earth and Mercury osculating orbital elements were retrieved from the WEB interface HORIZONS maintained by the NASA JPL. For the Sun's angular momentum, source of its post-Newtonian gravitomagnetic field, and quadrupole mass moment the values  \citep{1998MNRAS.297L..76P, 2017NSTIM.108.....V} $S_\odot=190.0\times 10^{39}~\textrm{kg~m}^2~\textrm{s}^{-1},~J_2^\odot=2.295\times 10^{-7}$ were adopted. The right ascension (RA) and declination (DEC) of the Sun's spin axis, referred to the Earth's mean equator at the epoch J2000.0,  are \citep{2007CeMDA..98..155S}
$\alpha_\odot = 286.13~\textrm{deg},~\delta_\odot = 63.87~\textrm{deg}$.}\label{Fig1}
\end{figure*}

\begin{figure*}
\centerline{
\begin{tabular}{c}
\epsfxsize= 12 cm\epsfbox{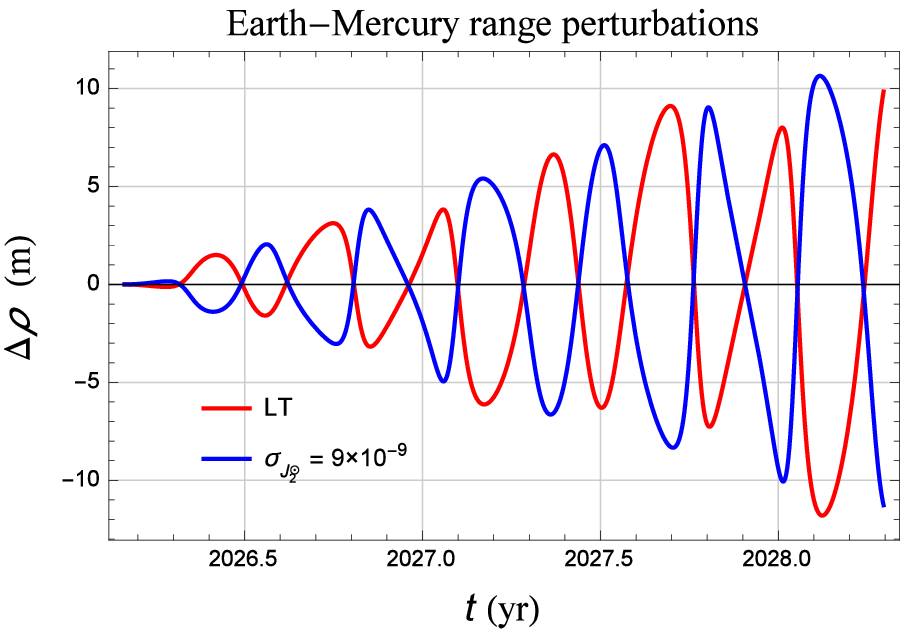}\\
\epsfxsize= 12 cm\epsfbox{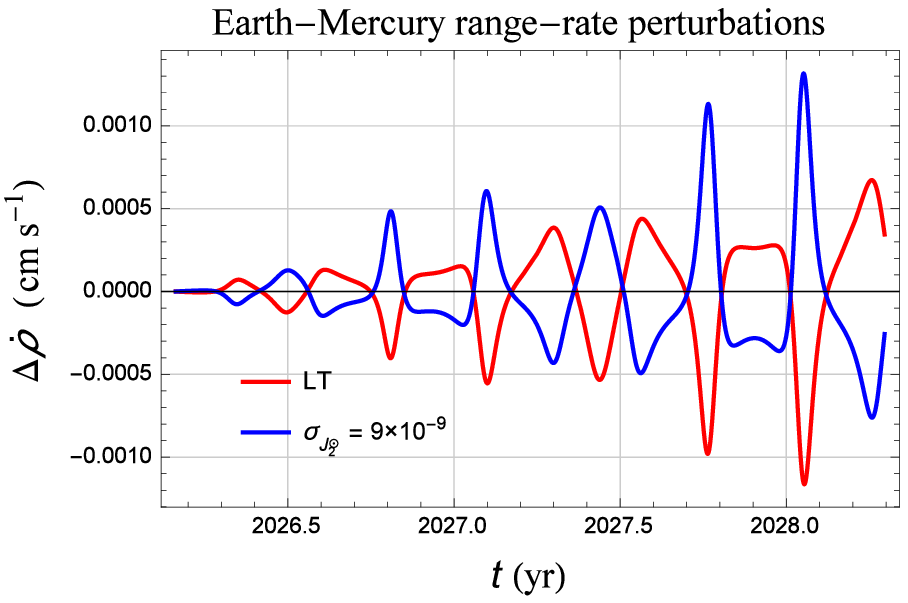}\\
\end{tabular}
}
\caption{Nominal Lense-Thirring and mismodeled $J_2^\odot$ perturbations of the Earth-Mercury range (in~\textrm{m}) and range-rate (in~$\textrm{cm~s}^{-1}$) during the expected extended mission of Bepi Colombo from 2026 March 14 to 2028 May 1. A coordinate system with the mean ecliptic at the epoch J2000.0 as fundamental reference $\grf{x,~y}$ plane was assumed. The initial values of the Earth and Mercury osculating orbital elements were retrieved from the WEB interface HORIZONS maintained by the NASA JPL. For the Sun's angular momentum, source of its post-Newtonian gravitomagnetic field, and the uncertainty in its quadrupole mass moment the values  \citep{1998MNRAS.297L..76P, 2017AJ....153..121P} $S_\odot=190.0\times 10^{39}~\textrm{kg~m}^2~\textrm{s}^{-1},~\sigma_{J_2^\odot}=9\times 10^{-9}$ were adopted. The right ascension (RA) and declination (DEC) of the Sun's spin axis, referred to the Earth's mean equator at the epoch J2000.0,  are \citep{2007CeMDA..98..155S}
$\alpha_\odot = 286.13~\textrm{deg},~\delta_\odot = 63.87~\textrm{deg}$.}\label{Fig2}
\end{figure*}

\begin{figure*}
\centerline{
\begin{tabular}{c}
\epsfxsize= 12 cm\epsfbox{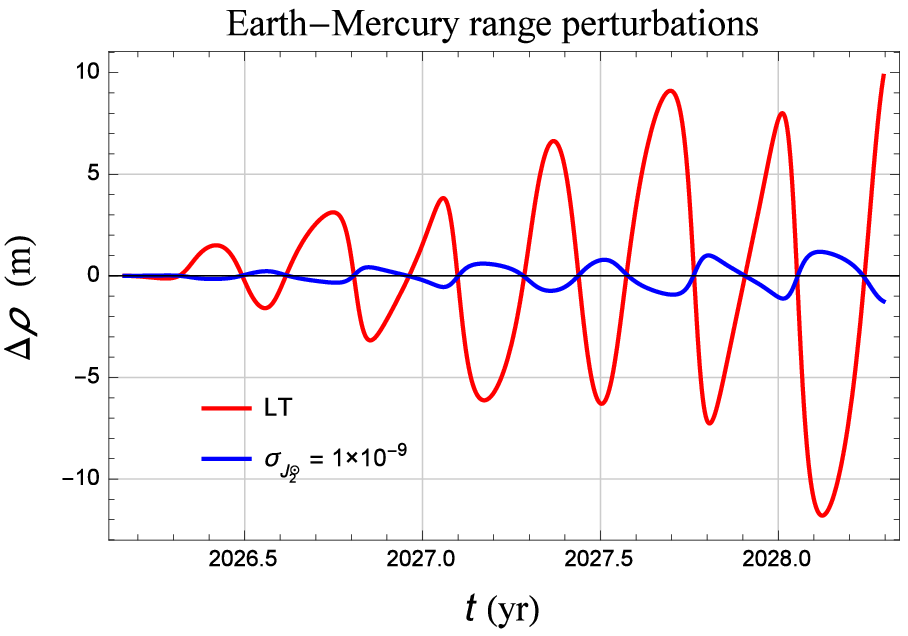}\\
\epsfxsize= 12 cm\epsfbox{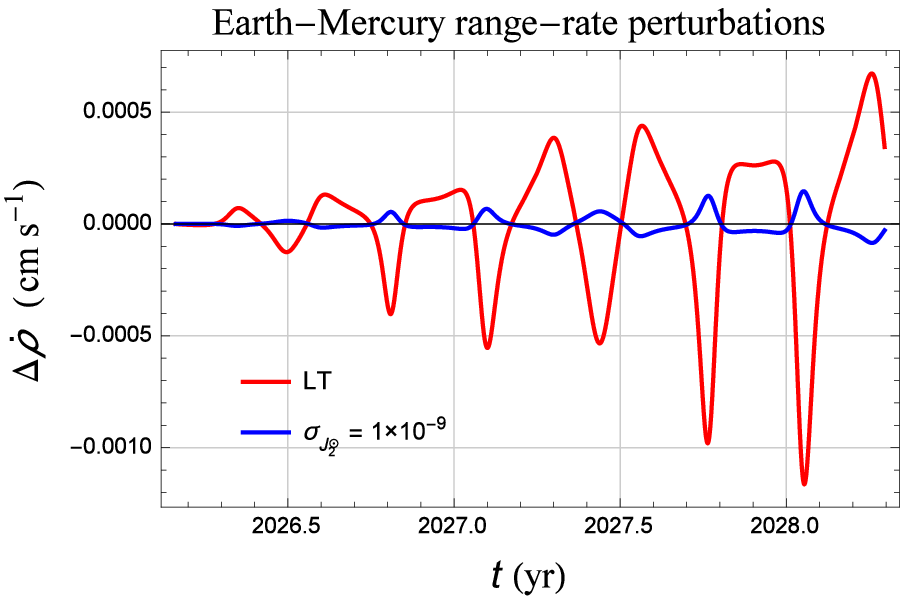}\\
\end{tabular}
}
\caption{Nominal Lense-Thirring and mismodeled $J_2^\odot$ perturbations of the Earth-Mercury range (in~\textrm{m}) and range-rate (in~$\textrm{cm~s}^{-1}$) during the expected extended mission of Bepi Colombo from 2026 March 14 to 2028 May 1. A coordinate system with the mean ecliptic at the epoch J2000.0 as fundamental reference $\grf{x,~y}$ plane was assumed. The initial values of the Earth and Mercury osculating orbital elements were retrieved from the WEB interface HORIZONS maintained by the NASA JPL. For the Sun's angular momentum, source of its post-Newtonian gravitomagnetic field, and the uncertainty in its quadrupole mass moment the values  \citep{1998MNRAS.297L..76P, 2017NSTIM.108.....V} $S_\odot=190.0\times 10^{39}~\textrm{kg~m}^2~\textrm{s}^{-1},~\sigma_{J_2^\odot}=1\times 10^{-9}$ were adopted. The right ascension (RA) and declination (DEC) of the Sun's spin axis, referred to the Earth's mean equator at the epoch J2000.0,  are \citep{2007CeMDA..98..155S}
$\alpha_\odot = 286.13~\textrm{deg},~\delta_\odot = 63.87~\textrm{deg}$.}\label{Fig3}
\end{figure*}

\begin{figure*}
\centerline{
\begin{tabular}{c}
\epsfxsize= 12 cm\epsfbox{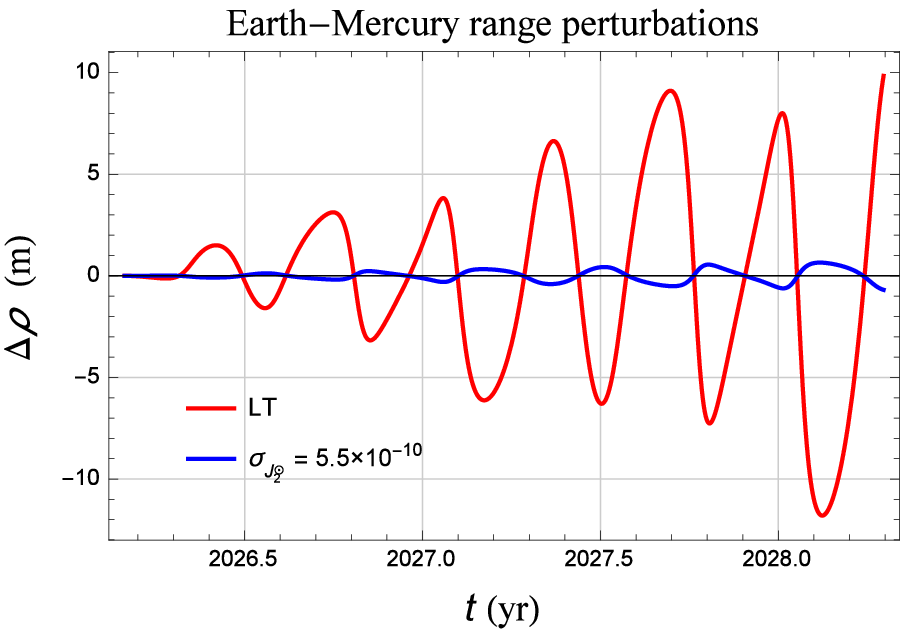}\\
\epsfxsize= 12 cm\epsfbox{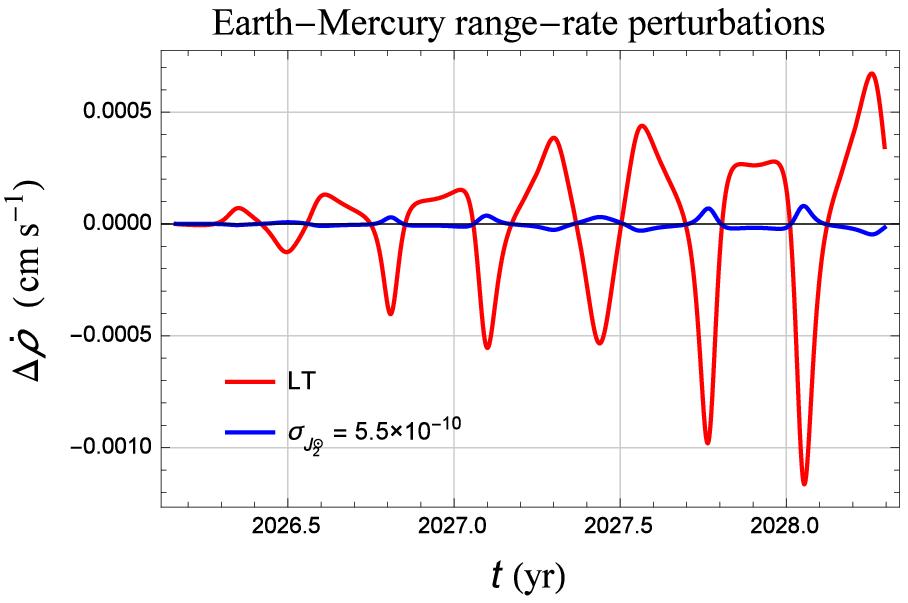}\\
\end{tabular}
}
\caption{Nominal Lense-Thirring and mismodeled $J_2^\odot$ perturbations of the Earth-Mercury range (in~\textrm{m}) and range-rate (in~$\textrm{cm~s}^{-1}$) during the expected extended mission of Bepi Colombo from 2026 March 14 to 2028 May 1. A coordinate system with the mean ecliptic at the epoch J2000.0 as fundamental reference $\grf{x,~y}$ plane was assumed. The initial values of the Earth and Mercury osculating orbital elements were retrieved from the WEB interface HORIZONS maintained by the NASA JPL. For the Sun's angular momentum, source of its post-Newtonian gravitomagnetic field, and the uncertainty in its quadrupole mass moment the values  \citep{1998MNRAS.297L..76P,2018Icar..301....9I} $S_\odot=190.0\times 10^{39}~\textrm{kg~m}^2~\textrm{s}^{-1},~\sigma_{J_2^\odot}=5.5\times 10^{-10}$ were adopted. The right ascension (RA) and declination (DEC) of the Sun's spin axis, referred to the Earth's mean equator at the epoch J2000.0,  are \citep{2007CeMDA..98..155S}
$\alpha_\odot = 286.13~\textrm{deg},~\delta_\odot = 63.87~\textrm{deg}$.}\label{Fig4}
\end{figure*}

In Figure~\ref{Fig1}, the nominal Earth-Mercury range and range-rate signatures induced by the Sun's angular momentum via the Lense-Thirring effect (red) and its quadrupole mass moment (blue) are depicted according to our analytical results in Sections~\ref{LTpert}~to~\ref{J2pert}.
In order to produce time series, we adopted the following expansion of the planetary true anomaly $f$ in terms of the mean anomaly $\mathcal{M}$
\citep[p.~77]{1961mcm..book.....B}
\begin{equation}
f\ton{t} = \mathcal{M}\ton{t} + 2\sum_{s = 1}^{s_\textrm{max}}\rp{1}{s}\grf{ J_s\ton{se} + \sum_{j = 1}^{j_\textrm{max}}\rp{\ton{1-\sqrt{1-e^2}}^j}{e^j}\qua{ J_{s-j}\ton{se} + J_{s+j}\ton{se}  }  }\sin s\mathcal{M}\ton{t}, \label{fMt}
\end{equation}
where $J_k\ton{se}$ is the Bessel function of the first kind of order $k$ and $s_\textrm{max},~j_\textrm{max}$ are some values of the summation indexes $s,~j$ set by the desired accuracy level.
It can be noted that the $J_2^\odot$-induced signals are much larger than the gravitomagnetic ones since their amplitudes can be as large as $300~\textrm{m}$ and $0.03~\textrm{cm~s}^{-1}$, while the general relativistic ones are as little as $10~\textrm{m}$ and $0.0010~\textrm{cm~s}^{-1}$. Such figures for the predicted Lense-Thirring range and range-rate shifts fall well within the previously mentioned improvements in the orbit determination of Mercury \citep{2010IAUS..261..356M}.
However, the Sun's quadrupole field is routinely included in the dynamical models of the current planetary ephemerides. Figure~\ref{Fig2} shows the mismodeled $J_2^\odot$ signature according to \citep{2017AJ....153..121P} $\sigma_{J_2^\odot}=9\times 10^{-9}$ recently inferred by processing the ranging data of the MESSENGER mission.
Now, the size of the quadrupolar signatures is at the same level of the gravitomagnetic ones; the picture highlights the different temporal patterns characterizing the two effects under consideration.
Finally, Figure~\ref{Fig3} depicts the case in which a much smaller uncertainty in the Sun's first even zonal is assumed: $\sigma_{J_2^\odot}=1\times 10^{-9}$ obtained with the recent INPOP17a ephemerides \citet{2017NSTIM.108.....V} in a global fit to an almost centennial record of data of several types including also, among other things, ranging to MESSENGER.
It turns out that the Lense-Thirring signatures are neatly predominant with respect to the residual traces left by the mismodeling in $J_2^\odot$ which are about 10 times smaller than the gravitomagnetic effects. Recently, \citet{2018NatureG} obtained, among other things, $\sigma_{J_2^\odot}=2.2\times 10^{-9}$ by analyzing the MESSENGER data.
It is important to remark that simulations of the BepiColombo mission, performed so far without modeling the Lense-Thirring effect, point towards an accuracy level in determining the Sun's quadrupole of the order of $\sigma_{J_2^\odot}\simeq 4.1-5.5\times 10^{-10}$ \citep{2016Univ....2...21S,2018Icar..301....9I}.
The corresponding plots are displayed in Figure~\ref{Fig4}; the maximum values of the mismodeled quadrupolar signals are now of the order of just $0.6~\textrm{m},~4\times 10^{-5}\textrm{cm~s}^{-1}$.

\begin{figure*}
\centerline{
\begin{tabular}{c}
\epsfxsize= 12 cm\epsfbox{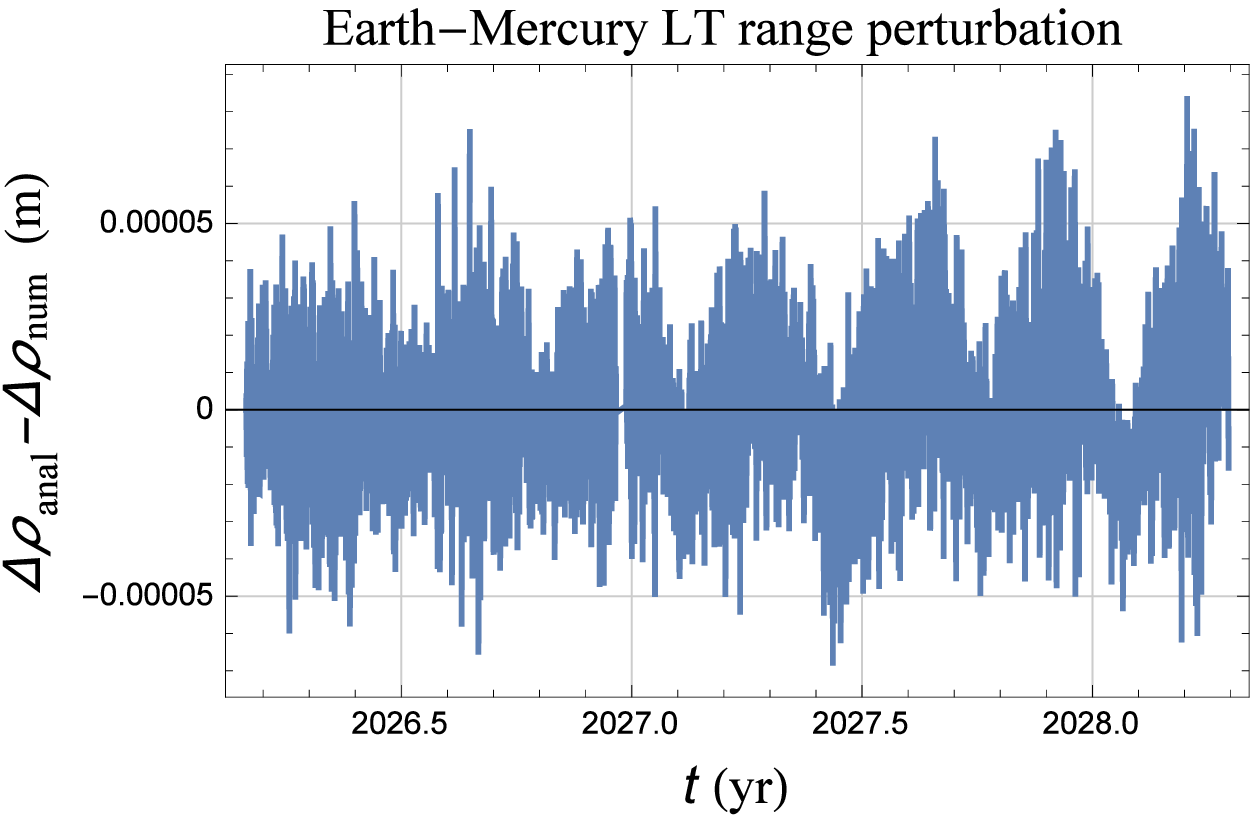}\\
\epsfxsize= 12 cm\epsfbox{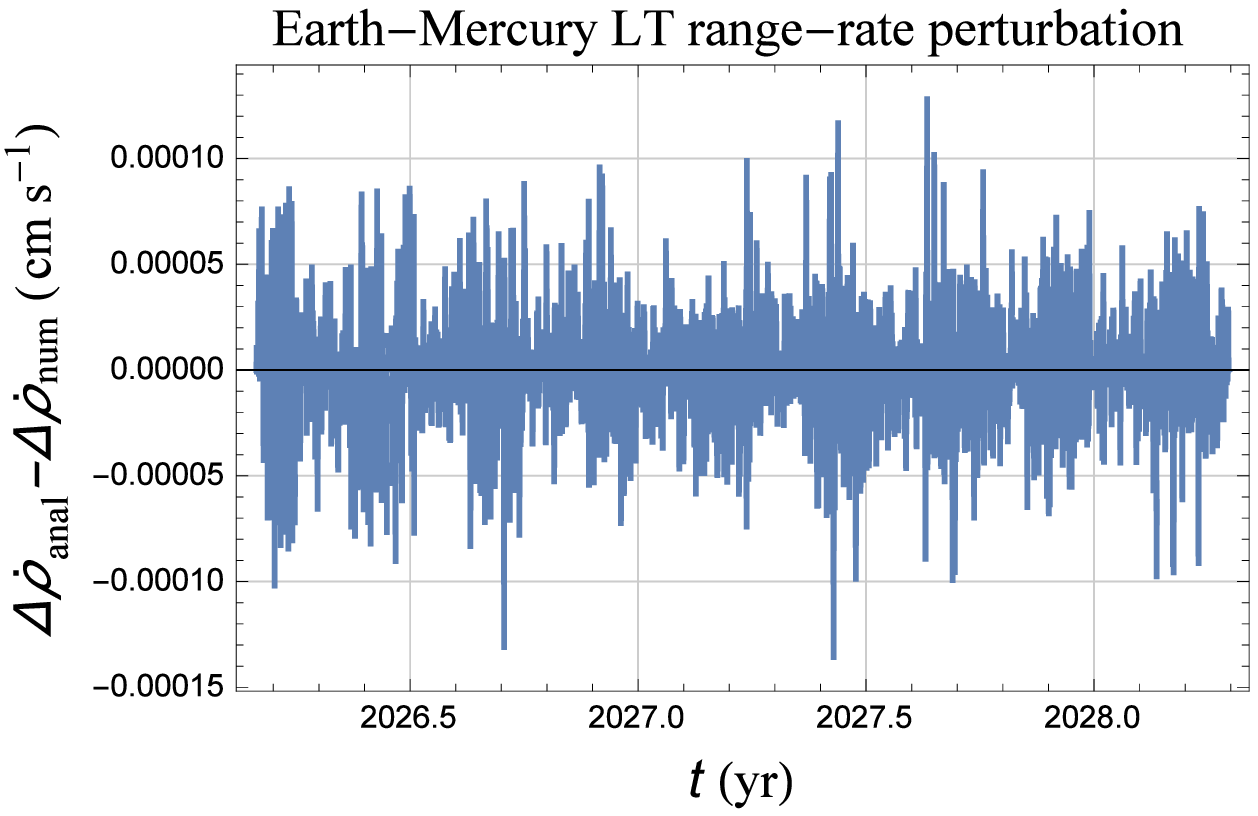}\\
\end{tabular}
}
\caption{Differences between the analytical and the numerical Lense-Thirring  perturbations of the Earth-Mercury range (in~\textrm{m}) and range-rate (in~$\textrm{cm~s}^{-1}$) during the expected extended mission of Bepi Colombo from 2026 March 14 to 2028 May 1. A coordinate system with the mean ecliptic at the epoch J2000.0 as fundamental reference $\grf{x,~y}$ plane was assumed. The initial values of the Earth and Mercury osculating orbital elements were retrieved from the WEB interface HORIZONS maintained by the NASA JPL. For the Sun's angular momentum, source of its post-Newtonian gravitomagnetic field,  the value  \citep{1998MNRAS.297L..76P} $S_\odot=190.0\times 10^{39}~\textrm{kg~m}^2~\textrm{s}^{-1}$ was adopted. The right ascension (RA) and declination (DEC) of the Sun's spin axis, referred to the Earth's mean equator at the epoch J2000.0,  are \citep{2007CeMDA..98..155S}
$\alpha_\odot = 286.13~\textrm{deg},~\delta_\odot = 63.87~\textrm{deg}$.}\label{Fig5bis}
\end{figure*}

Our analytical calculation is based on a perturbative approach in terms of the osculating Keplerian orbital elements, which are used in celestial mechanics, applied to the Lense-Thirring acceleration ${\bds A}_\textrm{LT}$ written in the standard harmonic post-Newtonian coordinates\footnote{According to http://iaaras.ru/en/dept/ephemeris/epm/2017/, ${\bds A}_\textrm{LT}$ was explicitly included in the latest version EPM2017 of the EPM ephemerides produced by the team led by E.V. Pitjeva.} \citep{1991ercm.book.....B,1989racm.book.....S,2010ITN....36....1P}. In order to check its validity, we numerically integrated the equations of motion of all the major bodies of the Solar System with and without the solar gravitomagnetic acceleration ${\bds A}_\textrm{LT}$ over the extended BepiColombo mission starting from the same initial conditions, and produced numerical time series $\Delta\rho_\textrm{num},~\Delta\dot\rho_\textrm{num}$ for the gravitomagnetic Earth-Mercury range and range-rate shifts of orbital origin. Then, we compared them to the corresponding analytically worked out time series $\Delta\rho_\textrm{anal},~\Delta\dot\rho_\textrm{anal}$ displayed in Figures~\ref{Fig1}~to~\ref{Fig4}. Figure~\ref{Fig5bis} depicts their differences; they are well below the expected experimental accuracy level since they amount to about $\left|\Delta\rho_\textrm{anal}-\Delta\rho_\textrm{num}\right|\lesssim 5\times 10^{-5}~\textrm{m},~\left|\Delta\dot\rho_\textrm{anal}-\Delta\dot\rho_\textrm{num}\right|\lesssim 1\times 10^{-4}~\textrm{cm~s}^{-1}$, respectively. Furthermore, Figure~\ref{Fig5bis} demonstrates pragmatically and effectively that using the osculating Keplerian orbital elements to perform our analytical gravitomagnetic calculation as in Section~\ref{LTpert} did not introduce any spurious, unphysical harmonics, contrary to what could, in principle, be argued on the basis of what happened in the literature for the post-Newtonian gravitoelectric field in the case of the two-body problem and its different orbital parameterizations \citep{1994ApJ...427..951K}. Moreover, to the best of the author's knowledge,  no other orbital parameterizations than the osculating elements have been used so far in the case of the Lense-Thirring effect; see, e.g., \citet{1988NCimB.101..127D} and references therein.

\begin{figure*}
\centerline{
\begin{tabular}{c}
\epsfxsize= 12 cm\epsfbox{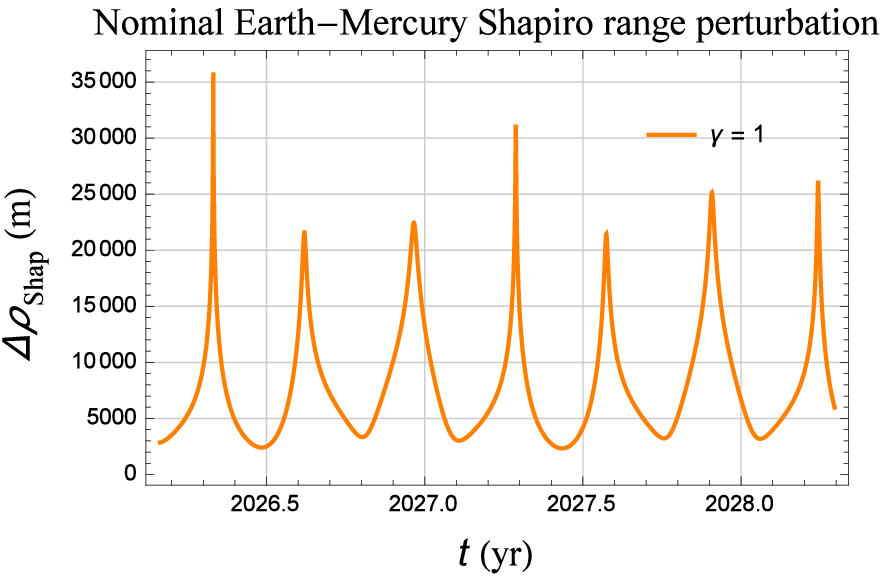}\\
\epsfxsize= 12 cm\epsfbox{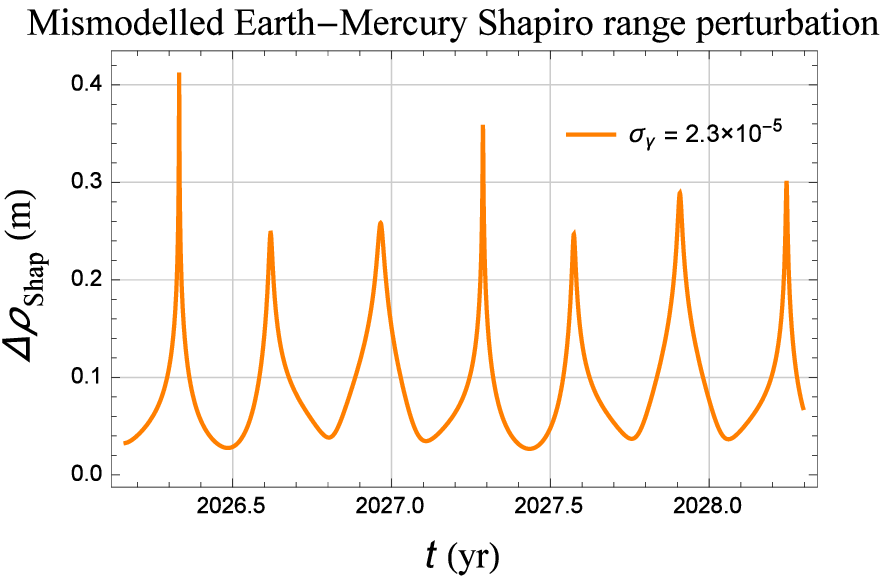}\\
\end{tabular}
}
\caption{Upper panel: nominal gravitoelectric range shift $\Delta\rho_\textrm{Shap}$ due to the standard Shapiro time delay $\Delta t_\textrm{Shap}$ for Mercury and the Earth over a time span corresponding to the currently planned extended mission of BepiColombo. The initial values of the Earth and Mercury state vectors were retrieved from the WEB interface HORIZONS maintained by the NASA JPL. Lower panel: mismodelled gravitoelectric range shift due to the Shapiro time delay for Mercury and the Earth over the same temporal interval as above by assuming $\sigma_\gamma = 2.3\times 10^{-5}$ \citep{2003Natur.425..374B} for the PPN parameter $\gamma$ entering the amplitude of such a propagation shift. Cfr. with the gravitomagnetic time series in Figures~\ref{Fig1}~to~\ref{Fig4}. }\label{Fig6bis}
\end{figure*}

As far as the gravitomagnetic propagation delay $\Delta t_\textrm{LT}$ is concerned, it turns out to be negligible in the present scenario. Indeed, it can be shown that is proportional to \citep{1997JMP....38.2587K, 1999ApJ...514..388W,2003PhLA..308..101C}
\eqi
\Delta t_\textrm{LT} \sim \rp{2GS}{c^4 r}\mathcal{F},\lb{dela}
\eqf
where $\mathcal{F}$ is a geometric factor depending on the mutual orientation of the primary's spin axis $\bds{\hat{S}}$ and the position vectors ${\mathbf{r}}_\textrm{A},~{\mathbf{r}}_\textrm{B}$ of the orbiting bodies.
For the Sun and Mercury, \rfr{dela} yields a time delay as little as $\left|\Delta t_\textrm{LT}\right| \lesssim 5\times 10^{-14}~\textrm{s}$, corresponding to a range shift of the order of just $\simeq 1.6\times 10^{-5}~\textrm{m}$. Furthermore, while the range perturbation of orbital origin is cumulative in time, the propagation delay of \rfr{dela} is periodic. The usual Shapiro time delay caused by the post-Newtonian gravitoelectric field of the Sun
\eqi
\Delta t_\textrm{Shap} = \rp{\ton{1+\gamma}\mu}{c^3}\ln\ton{\rp{r_\textrm{A}+r_\textrm{B}+\left|\mathbf{r}_\textrm{A}-\mathbf{r}_\textrm{B}\right|}{r_\textrm{A}+r_\textrm{B}-\left|\mathbf{r}_\textrm{A}-\mathbf{r}_\textrm{B}\right|}},
\eqf
where $\gamma$ is the parameter of the Parameterized Post-Newtonian (PPN) formalism accounting for the spatial curvature, yields an Earth-Mercury range signature which is nominally much larger than the gravitomagnetic one of orbital origin (see the upper panel of Figure~\ref{Fig6bis}). On the other hand, it is routinely included in the data reduction softwares, and its mismodeling due to the uncertainty in  $\gamma$ is small enough not to represent a problem, as shown by the lower panel of Figure~\ref{Fig6bis} obtained for $\sigma_\gamma = 2.3\times 10^{-5}$ \citep{2003Natur.425..374B}; indeed, it is apparent that it could not cancel out the Lense-Thirring Earth-Mercury range perturbation also because its temporal pattern is quite different.
In conclusion, it is not possible that the Lense-Thirring range shift, arising from the equations of motion of massive bodies, can be canceled by the range perturbations induced by the post-Newtonian delays in the propagation of the electromagnetic waves. Indeed, the latter ones have different temporal patterns, and their nominal (gravitomagnetic) and mismodelled (gravitoelectric) magnitudes are much smaller than the gravitomagnetic signal of interest.
\section{Gravitomagnetic corrections to the state vector of Mercury}
\begin{figure*}
\centerline{
\begin{tabular}{c}
\epsfxsize= 12 cm\epsfbox{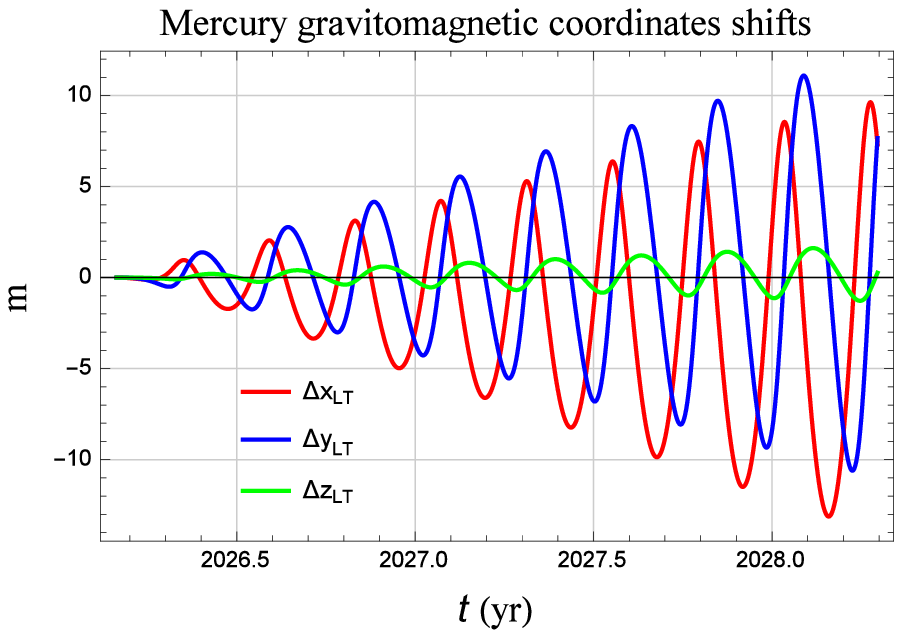}\\
\epsfxsize= 12 cm\epsfbox{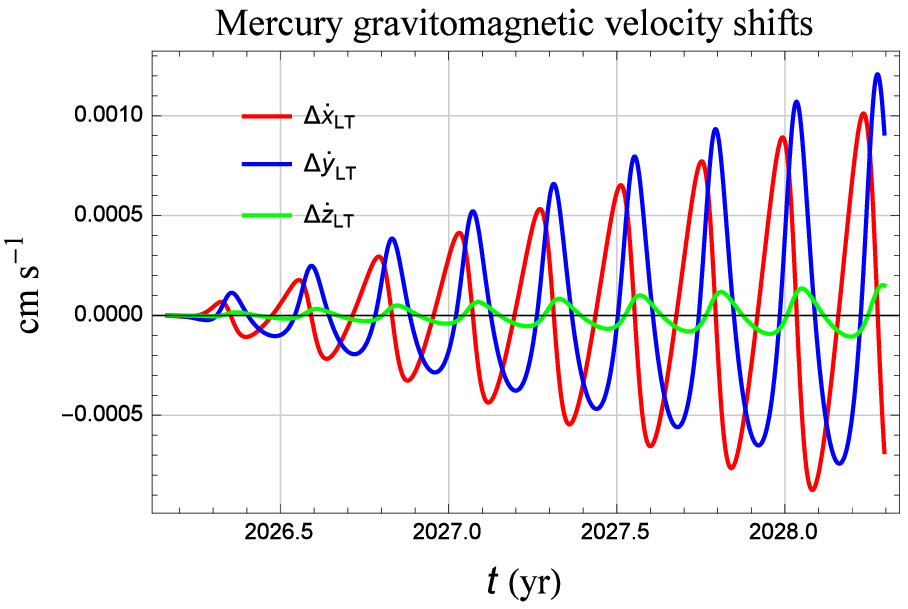}\\
\end{tabular}
}
\caption{Nominal Lense-Thirring corrections $\Delta x_\textrm{LT},~\Delta y_\textrm{LT},~\Delta z_\textrm{LT},~\Delta \dot x_\textrm{LT},~\Delta \dot y_\textrm{LT},~\Delta \dot z_\textrm{LT}$ to the heliocentric state vector of Mercury  during the expected extended mission of Bepi Colombo from 2026 March 14 to 2028 May 1. A coordinate system with the mean ecliptic at the epoch J2000.0 as fundamental reference $\grf{x,~y}$ plane was assumed. The initial values of the Hermean osculating orbital elements were retrieved from the WEB interface HORIZONS maintained by the NASA JPL. For the Sun's angular momentum, source of its post-Newtonian gravitomagnetic field,  the value  \citep{1998MNRAS.297L..76P} $S_\odot=190.0\times 10^{39}~\textrm{kg~m}^2~\textrm{s}^{-1}$ was adopted. The right ascension (RA) and declination (DEC) of the Sun's spin axis, referred to the Earth's mean equator at the epoch J2000.0,  are \citep{2007CeMDA..98..155S}
$\alpha_\odot = 286.13~\textrm{deg},~\delta_\odot = 63.87~\textrm{deg}$.}\label{Fig5}
\end{figure*}

\begin{figure*}
\centerline{
\begin{tabular}{c}
\epsfxsize= 12 cm\epsfbox{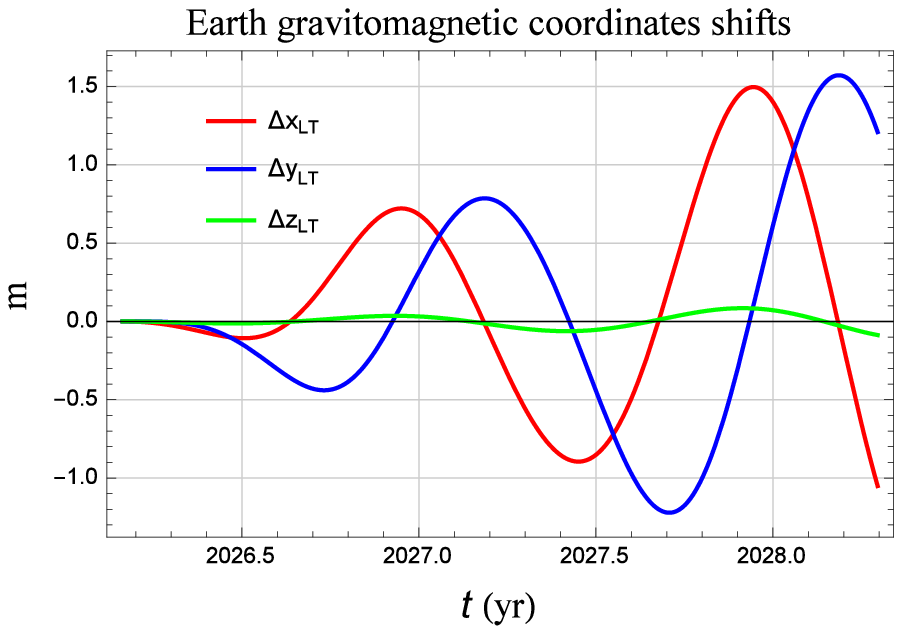}\\
\epsfxsize= 12 cm\epsfbox{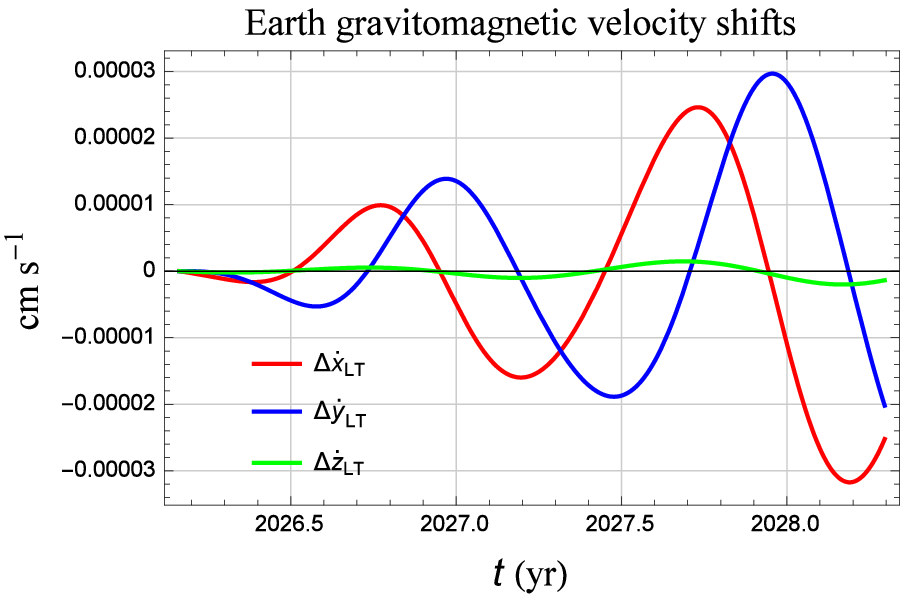}\\
\end{tabular}
}
\caption{Nominal Lense-Thirring corrections $\Delta x_\textrm{LT},~\Delta y_\textrm{LT},~\Delta z_\textrm{LT},~\Delta \dot x_\textrm{LT},~\Delta \dot y_\textrm{LT},~\Delta \dot z_\textrm{LT}$ to the heliocentric state vector of Earth  during the expected extended mission of Bepi Colombo from 2026 March 14 to 2028 May 1. A coordinate system with the mean ecliptic at the epoch J2000.0 as fundamental reference $\grf{x,~y}$ plane was assumed. The initial values of the terrestrial osculating orbital elements were retrieved from the WEB interface HORIZONS maintained by the NASA JPL. For the Sun's angular momentum, source of its post-Newtonian gravitomagnetic field,  the value  \citep{1998MNRAS.297L..76P} $S_\odot=190.0\times 10^{39}~\textrm{kg~m}^2~\textrm{s}^{-1}$ was adopted. The right ascension (RA) and declination (DEC) of the Sun's spin axis, referred to the Earth's mean equator at the epoch J2000.0,  are \citep{2007CeMDA..98..155S}
$\alpha_\odot = 286.13~\textrm{deg},~\delta_\odot = 63.87~\textrm{deg}$.}\label{Fig6}
\end{figure*}

\lb{caos}
In Figures~\ref{Fig5}~to~\ref{Fig6}, we plot the Lense-Thirring corrections $\Delta x_\textrm{LT},~\Delta y_\textrm{LT},~\Delta z_\textrm{LT},~\Delta \dot x_\textrm{LT},~\Delta \dot y_\textrm{LT},~\Delta \dot z_\textrm{LT}$ to the position and velocity vectors $\mathbf{r},~\mathbf{v}$ of Mercury and Earth  over a 2-yr time span; for the sake of definiteness, we adopt the timeframe of the extended mission of BepiColombo.
It turns out that the components  of the position vectors are shifted by about $10~\textrm{m}$ (Mercury) and $1.5~\textrm{m}$ (Earth), while the velocity's components are changed by about $0.001~\textrm{cm~s}^{-1}$ (Mercury) and $0.00003~\textrm{cm~s}^{-1}$ (Earth).

Such results may be used as possible suggestions to consider also the solar gravitomagnetic field in future accurate long-term integrations of the solar system aimed to test its stability over the eons. Indeed, it would be interesting to quantitatively check if neglecting the Lense-Thirring effect may have some appreciable impact on the outcomes of those existing studies predicting non-zero probabilities of interplanetary collisions or ejections. To the post-Newtonian level, only the gravitostatic, Schwarzschild-like acceleration was included so far in the existing literature; see, e.g., the Appendix in \citet{2015ApJ...811....9Z}. \citep{2009Natur.459..817L,2011A&A...532A..89L,2011ApJ...739...31L,2012A&A...548A..43B,2015ApJ...799..120B,2015ApJ...798....8Z,2015ApJ...811....9Z,2017AJ....154..193Z}. In view of Figures~\ref{Fig5}~to~\ref{Fig6}, it does not seem unreasonable since, as pointed out in \citet{1989Natur.338..237L,2015ApJ...811....9Z}, a
difference in initial position of $1~\textrm{cm}$ grows to about $1~\textrm{au}$  after $90-150~\textrm{Myr}$. Furthermore, in the analysis by \citet{2015ApJ...811....9Z}, the Hermean initial radial distance was offset by $1.75~\textrm{mm}$ between every two adjacent orbits while the largest overall offset was $2.80~\textrm{m}$. Such a figure is smaller than the typical Lense-Thirring shift after just two years, as per Figure~\ref{Fig5}.
\section{Summary and conclusions}\lb{fine}
In view of their wide application in practical investigations about several astronomical and astrophysical scenarios of interest for tests of fundamental physics, we looked at the mutual range $\rho$ and range-rate $\dot\rho$ of two test particles $\textrm{A},~\textrm{B}$ orbiting a common rotating primary. We set up an approach to explicitly calculate the perturbations $\Delta\rho,~\Delta\dot\rho$ due to a generic disturbing acceleration $\bds A$ for which an explicit analytical expression is available. We applied our computational strategy to the post-Newtonian and Newtonian effects of lowest order induced by the rotation of the central body: the general relativistic Lense-Thirring field, generated by the angular momentum $\bds S$ of the primary, and the classical perturbation caused by the first even zonal harmonic $J_2$ of its non-spherical potential. Our analytical results, which are particularly cumbersome in the case of the source's oblateness, are completely general since they do not rely upon any a priori simplifying assumptions pertaining both the particles' orbital configurations and the orientation of the primary's symmetry axis. Thus, they can be applied to any system of intercommunicating probes designing dedicated missions, like, e.g., a GRACE-type tandem orbiting some giant planet of our solar system, performing sensitivity analyses, reinterpreting existing data, and looking also at the long-term dynamics of, say, the inner solar system. We looked also at the general relativistic Schwarzschild-like effect by calculating exactly the corresponding full instantaneous shift of the mean anomaly; the variations of the other orbital elements can be found in the existing literature.

As a practical application of our analytical calculation, we considered the geocentric range and range-rate of Mercury during the planned extended phase of the forthcoming BepiColombo mission, to be launched in late 2018, which, among other things, should notably improve the Hermean ephemerides. It turned out that the expected nominal Lense-Thirring perturbations can reach the $10~\textrm{m},~1\times 10^{-3}~\textrm{cm~s}^{-1}$ level, well within the tracking accuracy of BepiColombo which is of the order of $\sigma_\rho\simeq 0.1~\textrm{m},~\sigma_{\dot\rho}\simeq 2\times 10^{-4}~\textrm{cm~s}^{-1}$. The competing signatures induced by the Sun's quadrupole moment $J_2^\odot$, if modeled at the level of accuracy reached by the recent INPOP17a ephemerides, i.e. $\sigma_{J_2^\odot}=1\times 10^{-9}$, would be about 10 times smaller than the relativistic signals of interest. Furthermore, BepiColombo should be able to constrain $J_2^\odot$ down to the  $\sigma_{J_2\odot}\simeq 5\times 10^{-10}$ level.
We successfully checked our analytical results for the Lense-Thirring range and range-rate shifts by comparing them to numerically produced ones by integrating the equations of motion; indeed, their differences ar as little as $\left|\Delta\rho_\textrm{anal}-\Delta\rho_\textrm{num}\right|\lesssim 5\times 10^{-5}~\textrm{m},~\left|\Delta\dot\rho_\textrm{anal}-\Delta\dot\rho_\textrm{num}\right|\lesssim 1\times 10^{-4}~\textrm{cm~s}^{-1}$. It turned out that the gravitomagnetic time delay due to the propagation of the electromagnetic waves is negligible since it leads to a range shift of the order of $\lesssim 10^{-5}~\textrm{m}$.
The gravitomagnetic field of the Sun has been always neglected so far in all the existing studies dedicated to the scientific return of BepiColombo. Our results show that it is time to explicitly account for the Lense-Thirring effect in future analyses in order to investigate its actual detectability.

We also looked at the Lense-Thirring corrections $\Delta x_\textrm{LT},~\Delta y_\textrm{LT},~\Delta z_\textrm{LT},~\Delta \dot x_\textrm{LT},~\Delta \dot y_\textrm{LT},~\Delta \dot z_\textrm{LT}$ to the state vectors of Mercury and Earth as a preliminary insight for future, accurate investigations of their impact on long-term integrations of the solar system dynamics over the past and future $\sim\textrm{Gyr}$ scale. We found that, after just two years, the position $\mathbf{r}$ and the velocity $\mathbf{v}$ of Mercury and Earth are changed by $10~\textrm{m},~1.5~\textrm{m}$ and $10^{-3}~\textrm{cm~s}^{-1},~10^{-5}~\textrm{cm~s}^{-1}$, respectively. In light of the existing studies, all neglecting the general relativistic gravitomagnetic field of the Sun, such shifts may not be negligible over the eons; suffice it to say that it has been demonstrated in the literature that an error as little as $1~\textrm{cm}$ grows to about $1~\textrm{au}$ after about $100~\textrm{Myr}$.
\section*{Acknowledgements}
I am grateful to an anonymous referee for his insightful comments and critical remarks
\bibliography{PXbib,IorioFupeng}{}

\begin{thebibliography}{65}
\expandafter\ifx\csname natexlab\endcsname\relax\def\natexlab#1{#1}\fi

\bibitem[{{Amaro-Seoane} {et~al}\mbox{.}(2013){Amaro-Seoane}, {Aoudia},
  {Babak}, {Bin{\'e}truy}, {Berti}, {Boh{\'e}}, {Caprini}, {Colpi}, {Cornish},
  {Danzmann}, {Dufaux}, {Gair}, {Hinder}, {Jennrich}, {Jetzer}, {Klein},
  {Lang}, {Lobo}, {Littenberg}, {McWilliams}, {Nelemans}, {Petiteau}, {Porter},
  {Schutz}, {Sesana}, {Stebbins}, {Sumner}, {Vallisneri}, {Vitale},
  {Volonteri}, {Ward}, \& {Wardell}}]{2013GWN.....6....4A}
{Amaro-Seoane} P. {et~al.}, 2013, GW Notes, 6, 4

\bibitem[{{Ashby}, {Bender} \& {Wahr}(2007){Ashby}, {Bender}, \&
  {Wahr}}]{2007PhRvD..75b2001A}
{Ashby} N., {Bender} P.~L., {Wahr} J.~M., 2007, Phys. Rev. D, 75, 022001

\bibitem[{{Balogh} {et~al}\mbox{.}(2007){Balogh}, {Grard}, {Solomon}, {Schulz},
  {Langevin}, {Kasaba}, \& {Fujimoto}}]{2007SSRv..132..611B}
{Balogh} A., {Grard} R., {Solomon} S.~C., {Schulz} R., {Langevin} Y., {Kasaba}
  Y., {Fujimoto} M., 2007, Space Sci. Rev., 132, 611

\bibitem[{{Batygin}, {Morbidelli} \& {Holman}(2015){Batygin}, {Morbidelli}, \&
  {Holman}}]{2015ApJ...799..120B}
{Batygin} K., {Morbidelli} A., {Holman} M.~J., 2015, ApJ, 799, 120

\bibitem[{{Bender} {et~al}\mbox{.}(2003){Bender}, {Hall}, {Ye}, \&
  {Klipstein}}]{2003SSRv..108..377B}
{Bender} P.~L., {Hall} J.~L., {Ye} J., {Klipstein} W.~M., 2003, Space Sci.
  Rev., 108, 377

\bibitem[{{Benkhoff} {et~al}\mbox{.}(2017){Benkhoff}, {Fujimoto}, {Murakami},
  \& {Zender}}]{2017EPSC...11..508B}
{Benkhoff} J., {Fujimoto} M., {Murakami} G., {Zender} J., 2017, European
  Planetary Science Congress, 11, EPSC2017

\bibitem[{{Benkhoff} {et~al}\mbox{.}(2010){Benkhoff}, {van Casteren},
  {Hayakawa}, {Fujimoto}, {Laakso}, {Novara}, {Ferri}, {Middleton}, \&
  {Ziethe}}]{2010P&SS...58....2B}
{Benkhoff} J. {et~al.}, 2010, Planet. Space Sci., 58, 2

\bibitem[{{Bertotti}, {Farinella} \& {Vokrouhlick\'{y}}(2003){Bertotti},
  {Farinella}, \& {Vokrouhlick\'{y}}}]{2003ASSL..293.....B}
{Bertotti} B., {Farinella} P., {Vokrouhlick\'{y}} D., 2003, {Physics of the
  Solar System - Dynamics and Evolution, Space Physics, and Spacetime
  Structure.} Kluwer, Dordrecht

\bibitem[{{Bertotti}, {Iess} \& {Tortora}(2003){Bertotti}, {Iess}, \&
  {Tortora}}]{2003Natur.425..374B}
{Bertotti} B., {Iess} L., {Tortora} P., 2003, Nature, 425, 374

\bibitem[{{Bou{\'e}}, {Laskar} \& {Farago}(2012){Bou{\'e}}, {Laskar}, \&
  {Farago}}]{2012A&A...548A..43B}
{Bou{\'e}} G., {Laskar} J., {Farago} F., 2012, A\&A, 548, A43

\bibitem[{{Brouwer} \& {Clemence}(1961)}]{1961mcm..book.....B}
{Brouwer} D., {Clemence} G.~M., 1961, {Methods of Celestial Mechanics}.
  Academic Press, New York

\bibitem[{{Brumberg}(1991)}]{1991ercm.book.....B}
{Brumberg} V.~A., 1991, {Essential Relativistic Celestial Mechanics}. Adam
  Hilger, Bristol

\bibitem[{{Brumberg}(2010{\natexlab{a}})}]{2010SchpJ...510669B}
{Brumberg} V.~A., 2010{\natexlab{a}}, Scholarpedia, 5

\bibitem[{{Brumberg}(2010{\natexlab{b}})}]{2010CeMDA.106..209B}
{Brumberg} V.~A., 2010{\natexlab{b}}, Celest. Mech. Dyn. Astr., 106, 209

\bibitem[{{Brumberg}(2013)}]{2013SoSyR..47..347B}
{Brumberg} V.~A., 2013, Solar Syst. Res., 47, 347

\bibitem[{{Canup} \& {Asphaug}(2001)}]{2001Natur.412..708C}
{Canup} R.~M., {Asphaug} E., 2001, Nature, 412, 708

\bibitem[{{Casotto}(1993)}]{1993CeMDA..55..209C}
{Casotto} S., 1993, Celest. Mech. Dyn. Astr., 55, 209

\bibitem[{{Cheng}(2002)}]{2002JGeod..76..169C}
{Cheng} M.~K., 2002, J. Geod., 76, 169

\bibitem[{{Ciufolini} {et~al}\mbox{.}(2003){Ciufolini}, {Kopeikin}, {Mashhoon},
  \& {Ricci}}]{2003PhLA..308..101C}
{Ciufolini} I., {Kopeikin} S., {Mashhoon} B., {Ricci} F., 2003, Phys. Lett. A,
  308, 101

\bibitem[{{Ciufolini} {et~al}\mbox{.}(2013){Ciufolini}, {Paolozzi}, {Koenig},
  {Pavlis}, {Ries}, {Matzner}, {Gurzadyan}, {Penrose}, {Sindoni}, \&
  {Paris}}]{2013NuPhS.243..180C}
{Ciufolini} I. {et~al.}, 2013, Nucl. Phys. B Proc. Suppl., 243, 180

\bibitem[{{Damour} \& {Schafer}(1988)}]{1988NCimB.101..127D}
{Damour} T., {Schafer} G., 1988, Nuovo Cimento B, 101, 127

\bibitem[{{Debono} \& {Smoot}(2016)}]{2016Univ....2...23D}
{Debono} I., {Smoot} G.~F., 2016, Universe, 2, 23

\bibitem[{{Everitt} {et~al}\mbox{.}(2011){Everitt}, {Debra}, {Parkinson},
  {Turneaure}, {Conklin}, {Heifetz}, {Keiser}, {Silbergleit}, {Holmes},
  {Kolodziejczak}, {Al-Meshari}, {Mester}, {Muhlfelder}, {Solomonik}, {Stahl},
  {Worden}, {Bencze}, {Buchman}, {Clarke}, {Al-Jadaan}, {Al-Jibreen}, {Li},
  {Lipa}, {Lockhart}, {Al-Suwaidan}, {Taber}, \& {Wang}}]{2011PhRvL.106v1101E}
{Everitt} C.~W.~F. {et~al.}, 2011, Phys. Rev. Lett., 106, 221101

\bibitem[{{Folkner}, {Jacobson} \& {Jones}(2015){Folkner}, {Jacobson}, \&
  {Jones}}]{2015IAUGA..2244873F}
{Folkner} W.~M., {Jacobson} R.~A., {Jones} D., 2015, IAU General Assembly, 22,
  2244873

\bibitem[{{Folkner} {et~al}\mbox{.}(2014){Folkner}, {Williams}, {Boggs},
  {Park}, \& {Kuchynka}}]{2014IPNPR.196C...1F}
{Folkner} W.~M., {Williams} J.~G., {Boggs} D.~H., {Park} R.~S., {Kuchynka} P.,
  2014, Interplanetary Network Progress Report, 196, 1

\bibitem[{{Genova} {et~al}\mbox{.}(2018){Genova}, {Mazarico}, {Goossens},
  {Lemoine}, {Neumann}, {Smith}, \& {Zuber}}]{2018NatureG}
{Genova} A., {Mazarico} E., {Goossens} S., {Lemoine} F.~G., {Neumann} G.~A.,
  {Smith} D.~E., {Zuber} M.~T., 2018, Nature Communications, 9, 289

\bibitem[{{Iess}, {Asmar} \& {Tortora}(2009){Iess}, {Asmar}, \&
  {Tortora}}]{2009AcAau..65..666I}
{Iess} L., {Asmar} S., {Tortora} P., 2009, Acta Astronautica, 65, 666

\bibitem[{{Imperi}, {Iess} \& {Mariani}(2018){Imperi}, {Iess}, \&
  {Mariani}}]{2018Icar..301....9I}
{Imperi} L., {Iess} L., {Mariani} M.~J., 2018, Icarus, 301, 9

\bibitem[{{Iorio}(2017)}]{2017EPJC...77..439I}
{Iorio} L., 2017, Eur. Phys. J. C, 77, 439

\bibitem[{{Iorio} {et~al}\mbox{.}(2011){Iorio}, {Lichtenegger}, {Ruggiero}, \&
  {Corda}}]{2011Ap&SS.331..351I}
{Iorio} L., {Lichtenegger} H.~I.~M., {Ruggiero} M.~L., {Corda} C., 2011,
  Astrophys. Space Sci., 331, 351

\bibitem[{{Kim} \& {Lee}(2009)}]{2009AcAau..65.1571K}
{Kim} J., {Lee} S.~W., 2009, Acta Astronaut., 65, 1571

\bibitem[{{Klioner} \& {Kopeikin}(1994)}]{1994ApJ...427..951K}
{Klioner} S.~A., {Kopeikin} S.~M., 1994, ApJ, 427, 951

\bibitem[{{Konopliv} {et~al}\mbox{.}(2011){Konopliv}, {Asmar}, {Folkner},
  {Karatekin}, {Nunes}, {Smrekar}, {Yoder}, \& {Zuber}}]{2011Icar..211..401K}
{Konopliv} A.~S., {Asmar} S.~W., {Folkner} W.~M., {Karatekin} {\"O}., {Nunes}
  D.~C., {Smrekar} S.~E., {Yoder} C.~F., {Zuber} M.~T., 2011, Icarus, 211, 401

\bibitem[{{Kopeikin}, {Efroimsky} \& {Kaplan}(2011){Kopeikin}, {Efroimsky}, \&
  {Kaplan}}]{2011rcms.book.....K}
{Kopeikin} S., {Efroimsky} M., {Kaplan} G., 2011, {Relativistic Celestial
  Mechanics of the Solar System}. Wiley-VCH, Weinheim

\bibitem[{{Kopeikin}(1997)}]{1997JMP....38.2587K}
{Kopeikin} S.~M., 1997, J. Math. Phys., 38, 2587

\bibitem[{{Laskar}(1989)}]{1989Natur.338..237L}
{Laskar} J., 1989, Nature, 338, 237

\bibitem[{{Laskar} {et~al}\mbox{.}(2011){Laskar}, {Fienga}, {Gastineau}, \&
  {Manche}}]{2011A&A...532A..89L}
{Laskar} J., {Fienga} A., {Gastineau} M., {Manche} H., 2011, A\&A, 532, A89

\bibitem[{{Laskar} \& {Gastineau}(2009)}]{2009Natur.459..817L}
{Laskar} J., {Gastineau} M., 2009, Nature, 459, 817

\bibitem[{{Lithwick} \& {Wu}(2011)}]{2011ApJ...739...31L}
{Lithwick} Y., {Wu} Y., 2011, ApJ, 739, 31

\bibitem[{{Loomis}, {Nerem} \& {Luthcke}(2012){Loomis}, {Nerem}, \&
  {Luthcke}}]{2012JGeod..86..319L}
{Loomis} B.~D., {Nerem} R.~S., {Luthcke} S.~B., 2012, J. Geod., 86, 319

\bibitem[{{Milani}, {Nobili} \& {Farinella}(1987){Milani}, {Nobili}, \&
  {Farinella}}]{Nobilibook87}
{Milani} A., {Nobili} A., {Farinella} P., 1987, {Non-gravitational
  perturbations and satellite geodesy}. Adam Hilger, Bristol

\bibitem[{{Milani} {et~al}\mbox{.}(2010){Milani}, {Tommei}, {Vokrouhlick{\'y}},
  {Latorre}, \& {Cical{\`o}}}]{2010IAUS..261..356M}
{Milani} A., {Tommei} G., {Vokrouhlick{\'y}} D., {Latorre} E., {Cical{\`o}} S.,
  2010, in IAU Symposium, Vol. 261, Relativity in Fundamental Astronomy:
  Dynamics, Reference Frames, and Data Analysis, {Klioner} S.~A., {Seidelmann}
  P.~K., {Soffel} M.~H., eds., Cambridge University Press, Cambridge, pp.
  356--365

\bibitem[{{Milani} {et~al}\mbox{.}(2002){Milani}, {Vokrouhlick{\'y}},
  {Villani}, {Bonanno}, \& {Rossi}}]{2002PhRvD..66h2001M}
{Milani} A., {Vokrouhlick{\'y}} D., {Villani} D., {Bonanno} C., {Rossi} A.,
  2002, Phys. Rev. D, 66, 082001

\bibitem[{{Nobili} \& {Will}(1986)}]{1986Natur.320...39N}
{Nobili} A.~M., {Will} C.~M., 1986, Nature, 320, 39

\bibitem[{{Park}, {Folkner} \& {Konopliv}(2015){Park}, {Folkner}, \&
  {Konopliv}}]{2015IAUGA..2227771P}
{Park} R.~S., {Folkner} W.~M., {Konopliv} A.~S., 2015, IAU General Assembly,
  22, 2227771

\bibitem[{{Park} {et~al}\mbox{.}(2017){Park}, {Folkner}, {Konopliv},
  {Williams}, {Smith}, \& {Zuber}}]{2017AJ....153..121P}
{Park} R.~S., {Folkner} W.~M., {Konopliv} A.~S., {Williams} J.~G., {Smith}
  D.~E., {Zuber} M.~T., 2017, AJ, 153, 121

\bibitem[{{Petit} \& {Luzum}(2010)}]{2010ITN....36....1P}
{Petit} G., {Luzum} B., 2010, IERS Technical Note, 36, p. 156

\bibitem[{{Pijpers}(1998)}]{1998MNRAS.297L..76P}
{Pijpers} F.~P., 1998, MNRAS, 297, L76

\bibitem[{{Renzetti}(2013)}]{2013CEJPh..11..531R}
{Renzetti} G., 2013, Centr. Eur. J. Phys., 11, 531

\bibitem[{{Schettino} \& {Tommei}(2016)}]{2016Univ....2...21S}
{Schettino} G., {Tommei} G., 2016, Universe, 2, 21

\bibitem[{{Seidelmann} {et~al}\mbox{.}(2007){Seidelmann}, {Archinal},
  {A'Hearn}, {Conrad}, {Consolmagno}, {Hestroffer}, {Hilton}, {Krasinsky},
  {Neumann}, {Oberst}, {Stooke}, {Tedesco}, {Tholen}, {Thomas}, \&
  {Williams}}]{2007CeMDA..98..155S}
{Seidelmann} P.~K. {et~al.}, 2007, Celest. Mech. Dyn. Astr., 98, 155

\bibitem[{{Sheard} {et~al}\mbox{.}(2012){Sheard}, {Heinzel}, {Danzmann},
  {Shaddock}, {Klipstein}, \& {Folkner}}]{2012JGeod..86.1083S}
{Sheard} B.~S., {Heinzel} G., {Danzmann} K., {Shaddock} D.~A., {Klipstein}
  W.~M., {Folkner} W.~M., 2012, J. Geod., 86, 1083

\bibitem[{{Soffel} {et~al}\mbox{.}(2003){Soffel}, {Klioner}, {Petit}, {Wolf},
  {Kopeikin}, {Bretagnon}, {Brumberg}, {Capitaine}, {Damour}, {Fukushima},
  {Guinot}, {Huang}, {Lindegren}, {Ma}, {Nordtvedt}, {Ries}, {Seidelmann},
  {Vokrouhlick{\'y}}, {Will}, \& {Xu}}]{2003AJ....126.2687S}
{Soffel} M. {et~al.}, 2003, AJ, 126, 2687

\bibitem[{{Soffel}(1989)}]{1989racm.book.....S}
{Soffel} M.~H., 1989, {Relativity in Astrometry, Celestial Mechanics and
  Geodesy}. Springer-Verlag; Berlin Heidelberg New York

\bibitem[{{Tapley} {et~al}\mbox{.}(2004){Tapley}, {Bettadpur}, {Watkins}, \&
  {Reigber}}]{2004GeoRL..31.9607T}
{Tapley} B.~D., {Bettadpur} S., {Watkins} M., {Reigber} C., 2004, Geophys. Res.
  Lett., 31, L09607

\bibitem[{{Thorne}(1986)}]{1986hmac.book..103T}
{Thorne} K.~S., 1986, in Highlights of Modern Astrophysics: Concepts and
  Controversies, {Shapiro} S.~L., {Teukolsky} S.~A., {Salpeter} E.~E., eds.,
  Wiley, NY, p. 103

\bibitem[{{Thorne}(1988)}]{1988nznf.conf..573T}
{Thorne} K.~S., 1988, in Near Zero: New Frontiers of Physics, {Fairbank} J.~D.,
  {Deaver} Jr. B.~S., {Everitt} C.~W.~F., {Michelson} P.~F., eds., Freeman, New
  York, NY, pp. 573--586

\bibitem[{{Verma} {et~al}\mbox{.}(2014){Verma}, {Fienga}, {Laskar}, {Manche},
  \& {Gastineau}}]{2014A&A...561A.115V}
{Verma} A.~K., {Fienga} A., {Laskar} J., {Manche} H., {Gastineau} M., 2014,
  A\&A, 561, A115

\bibitem[{{Viswanathan} {et~al}\mbox{.}(2017){Viswanathan}, {Fienga},
  {Gastineau}, \& {Laskar}}]{2017NSTIM.108.....V}
{Viswanathan} V., {Fienga} A., {Gastineau} M., {Laskar} J., 2017, Notes
  Scientifiques et Techniques de l'Institut de M\'{e}canique C\'{e}leste, 108

\bibitem[{{Wex} \& {Kopeikin}(1999)}]{1999ApJ...514..388W}
{Wex} N., {Kopeikin} S.~M., 1999, ApJ, 514, 388

\bibitem[{{Wolff}(1969)}]{1969JGR....74.5295W}
{Wolff} M., 1969, J. Geophys. Res., 74, 5295

\bibitem[{{Zeebe}(2015{\natexlab{a}})}]{2015ApJ...798....8Z}
{Zeebe} R.~E., 2015{\natexlab{a}}, ApJ, 798, 8

\bibitem[{{Zeebe}(2015{\natexlab{b}})}]{2015ApJ...811....9Z}
{Zeebe} R.~E., 2015{\natexlab{b}}, ApJ, 811, 9

\bibitem[{{Zeebe}(2017)}]{2017AJ....154..193Z}
{Zeebe} R.~E., 2017, AJ, 154, 193

\bibitem[{{Zuber} {et~al}\mbox{.}(2013){Zuber}, {Smith}, {Lehman}, {Hoffman},
  {Asmar}, \& {Watkins}}]{2013SSRv..178....3Z}
{Zuber} M.~T., {Smith} D.~E., {Lehman} D.~H., {Hoffman} T.~L., {Asmar} S.~W.,
  {Watkins} M.~M., 2013, Space Sci. Rev., 178, 3

\end{thebibliography}
%-----------------------------------------

\end{document}